\documentclass[journal]{IEEEtran}
\pdfoutput=1
\ifCLASSINFOpdf
\else
\fi

\usepackage{cite}
\usepackage{amsmath,amssymb,amsfonts}
\usepackage{algorithmic}
\usepackage[ruled,vlined]{algorithm2e}
\usepackage{graphicx}
\usepackage{textcomp}
\usepackage{xcolor}
\def\BibTeX{{\rm B\kern-.05em{\sc i\kern-.025em b}\kern-.08em
    T\kern-.1667em\lower.7ex\hbox{E}\kern-.125emX}}

\usepackage{booktabs} 
\usepackage{multicol}
\usepackage{lipsum}
\usepackage{mwe}
\usepackage{tabularx}
\usepackage{multirow}
\usepackage{caption,subcaption}
\usepackage{cleveref}
\usepackage{color}
\usepackage{float}
\usepackage{authblk}
\usepackage{subfloat}
\usepackage{threeparttable}
\begin{document}
%
\title{DeepNVM++: Cross-Layer Modeling and Optimization Framework of Non-Volatile Memories for Deep Learning}


%
%
%

\author{Ahmet Inci, Mehmet Meric Isgenc, Diana~Marculescu,~\IEEEmembership{Fellow,~IEEE}
\thanks{Ahmet Inci is with the Department
of Electrical and Computer Engineering, Carnegie Mellon University, Pittsburgh,
PA, 15213 USA e-mail: ainci@andrew.cmu.edu}
\thanks{Mehmet Meric Isgenc was with the Department
of Electrical and Computer Engineering, Carnegie Mellon University, Pittsburgh,
PA, 15213 USA e-mail: misgenc@andrew.cmu.edu}
\thanks{Diana Marculescu is with the Department of Electrical and Computer Engineering, The University of Texas at Austin, Austin, TX, 78712 USA and Carnegie Mellon University, Pittsburgh, PA, 15213 USA e-mail: dianam@utexas.edu}
\thanks{Preprint}}

\maketitle

\begin{abstract}
Non-volatile memory (NVM) technologies such as spin-transfer torque magnetic random access memory (STT-MRAM) and spin-orbit torque magnetic random access memory (SOT-MRAM) have significant advantages compared to conventional SRAM due to their non-volatility, higher cell density, and scalability features. While previous work has investigated several architectural implications of NVM for generic applications, in this work we present \textit{DeepNVM++}, a \textit{framework} to characterize, model, and analyze NVM-based caches in GPU architectures for deep learning (DL) applications by combining technology-specific circuit-level models and the actual memory behavior of various DL workloads. We present both \textit{iso-capacity} and \textit{iso-area} performance and energy analysis for systems whose last-level caches rely on conventional SRAM and emerging STT-MRAM and SOT-MRAM technologies. In the iso-capacity case, STT-MRAM and SOT-MRAM provide up to $3.8 \times$ and $4.7 \times$ energy-delay product (EDP) reduction and $2.4 \times$ and $2.8 \times$ area reduction compared to conventional SRAM, respectively. Under iso-area assumptions, STT-MRAM and SOT-MRAM provide up to $2 \times$ and $2.3 \times$ EDP reduction and accommodate $2.3 \times$ and $3.3 \times$ cache capacity when compared to SRAM, respectively. We also perform a scalability analysis and show that STT-MRAM and SOT-MRAM achieve orders of magnitude EDP reduction when compared to SRAM for large cache capacities. Our comprehensive cross-layer framework is demonstrated on STT-/SOT-MRAM technologies and can be used for the characterization, modeling, and analysis of \textit{any} NVM technology for last-level caches in GPUs for DL applications.
\end{abstract}


%
\IEEEpeerreviewmaketitle

\section{Introduction} 
\IEEEPARstart{O}{ver} the last decade, the performance boost achieved through CMOS scaling has plateaued, necessitating sophisticated computer architecture solutions to gain higher performance in computing systems while maintaining a feasible power density. These objectives, however, are concurrently challenged by the limitations of the performance of memory resources \cite{Wulf:1995:HMW:216585.216588}. In contrast to the initial insight of Dennard on power density \cite{dennard}, deep CMOS scaling has exacerbated static power consumption, causing the heat density of ICs to reach catastrophic levels unless properly addressed \cite{temp_control,coskun2007temp,coskun2008temp}. 

As computers suffer from memory and power related limitations, the demand for data-intensive applications has been on the rise. With the increasing data deluge and recent improvements in GPU architectures, deep neural networks (DNNs) have achieved remarkable success in various tasks such as image recognition \cite{EfficientNet,FixRes}, object detection \cite{EfficientDet}, and chip placement \cite{ChipPlacement} by utilizing inherent massive parallelism of GPU platforms. However, DNN workloads continue to have large memory footprints and significant computational requirements to achieve higher accuracy. Thus, DNN workloads exacerbate the memory bottleneck which degrades the overall performance of the system. To this end, while DL practitioners focus on model compression techniques \cite{han2016deep,ruizhou2018lightnn,Chin2020CVPR}, hardware designers augment memory capacities of GPU architectures to overcome the memory bottleneck problem and improve the overall system performance. We note the current trend of GPU architectures is towards increasing last-level cache capacity as shown in Figure~\ref{fig:trend}. Our analysis shows that conventional SRAM technology incurs scalability problems as far as power, performance, and area (PPA) is concerned \cite{sram_technology,sram_leakage,nvm_llc_stt}. Non-volatile memory (NVM) technology is one of the most promising solutions to tackle memory bottleneck problem for data-intensive applications \cite{3dmram2008dac}.
However, because much of emerging NVM technology is not available for commercial use, there is an obvious need for a framework to perform design space exploration for these emerging NVM technologies for deep learning (DL) workloads.

\begin{figure}[!tbp]
  \centering
  \subfloat{\includegraphics[width=0.47\textwidth]{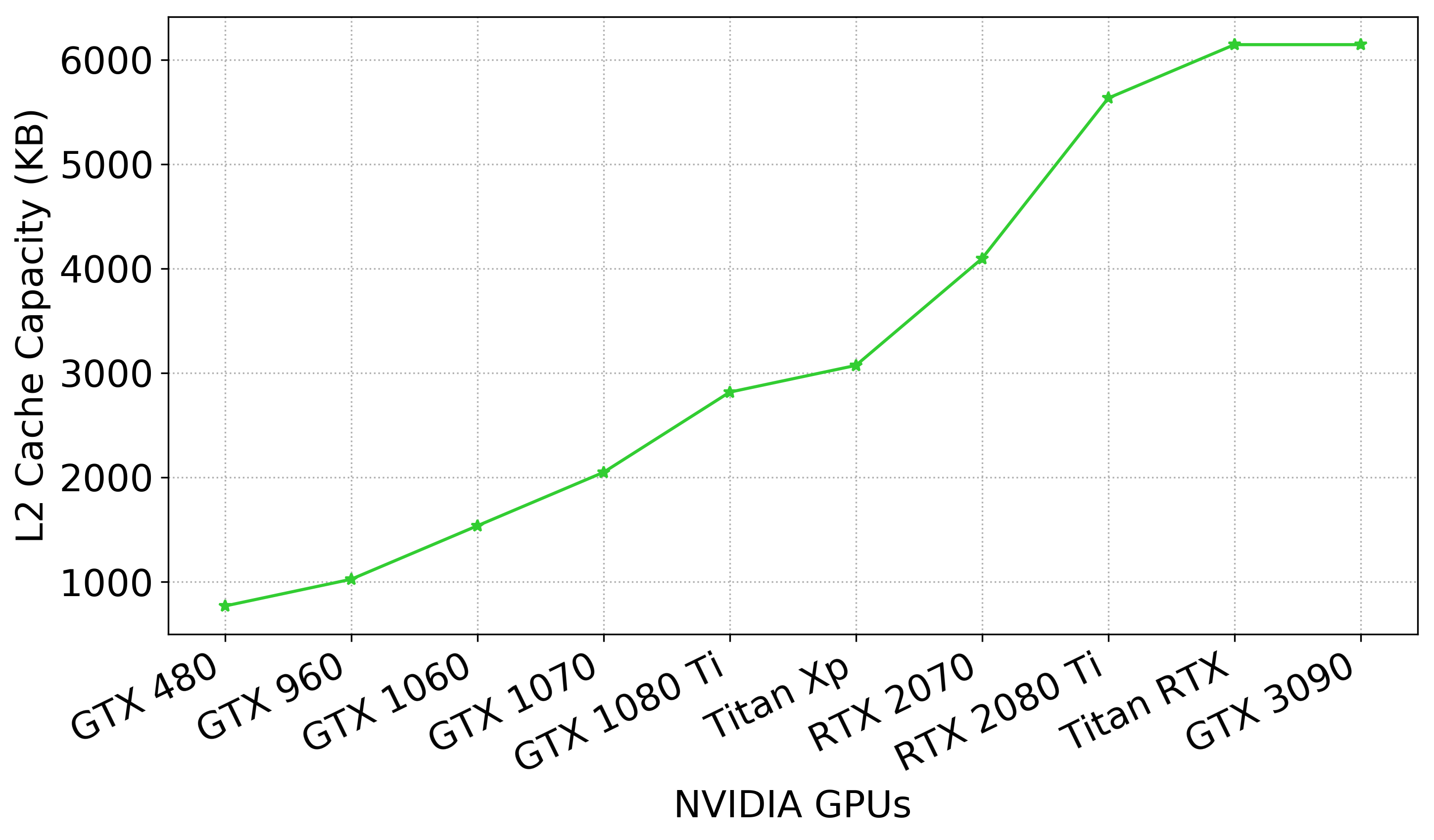}}
  \label{fig:trend}
 \caption{L2 cache capacity in recent NVIDIA GPUs~\cite{nvidia_trend}}
 \label{fig:trend}
\end{figure}

In this work, we present \textit{DeepNVM++}, an extended and improved framework \cite{DeepNVM} to characterize, model, and optimize NVM-based caches in GPU architectures for deep learning workloads. Without loss of generality, we demonstrate our framework for spin-transfer torque magnetic random access memory (STT-MRAM) and spin-orbit torque magnetic random access memory (SOT-MRAM), keeping in mind that it can be used for any NVM technology, GPU platform, or deep learning workload. Our cross-layer analysis framework incorporates both circuit-level characterization aspects and the memory behavior of various DL workloads running on an actual GPU platform. \textit{DeepNVM++} enables the evaluation of \textit{power, performance, and area} of NVMs when used for last-level (L2) caches in GPUs and seeks to exploit the benefits of this emerging technology to improve the performance of deep learning applications.

To perform \textit{iso-capacity} analysis, we carry out extensive memory profiling of various deep learning workloads for both training and inference on existing GPU platforms. For the \textit{iso-area} analysis, existing platforms cannot be used for varying cache sizes, so we rely on architecture-level simulation of GPUs to quantify and better understand last-level cache capacity and off-chip memory accesses. In both cases, our framework automatically combines resulting memory statistics with circuit and microarchitecture-level characterization and analysis of emerging NVM technologies to gauge their impact on DL workloads running on future GPU-based platforms.

\section{Related Work and Paper Contributions}
Although 16nm has become a commonplace technology for high-end customers of foundries, an intriguing inflection point awaits the electronics community as we approach the end of the traditional density, power, and performance benefits of CMOS scaling. To move beyond the computing limitations imposed by staggering CMOS scaling trends, MRAM has emerged as a promising candidate. 

The enabling technology of MRAM consists of magnetic tunnel junction (MTJ) pillars that can store data as a resistive state. An MTJ pillar consists of a thin oxide film sandwiched by two ferromagnetic layers. One of these ferromagnetic layers has a fixed magnetization which serves as a reference layer. The magnetization of the other layer can be altered by changing the direction of the current that flows through the pillar. If the magnetization of the free layer and the reference layer are in parallel, the device is in the low resistance state. If the magnetization of layers is in opposite directions, the device is in the high resistance state \cite{mtj}. 

STT bitcells \cite{stt} use an MTJ pillar as their core storage element and an additional access transistor to enable read and write operations. Although STT bitcells offer non-volatility, low read latency, and high endurance \cite{endurance}, the write current is also high \cite{stt_high_current,survey,yalamanchili2010stt}, which increases power consumption. To this end, SOT bitcells have been proposed to overcome the write current challenges by isolating the read and write paths \cite{mehdi2016sot}. Because the read disturbance errors are much less likely in SOT bitcells, both read and write access devices can be tuned in accordance with the lower current requirements \cite{mehdiNVM,oboril2015TCAD}. The read and write current requirements of STT and SOT bitcells can have a crucial impact on the eventual MRAM characteristics because they affect the CMOS access transistors, bitcell area, and peripheral logic. Thus, a comparison of these bitcells and the traditional SRAM merits a meticulous analysis that take these factors into account.

\begin{figure}[!tbp]
  \centering
 \subfloat{\includegraphics[width=0.5\textwidth]{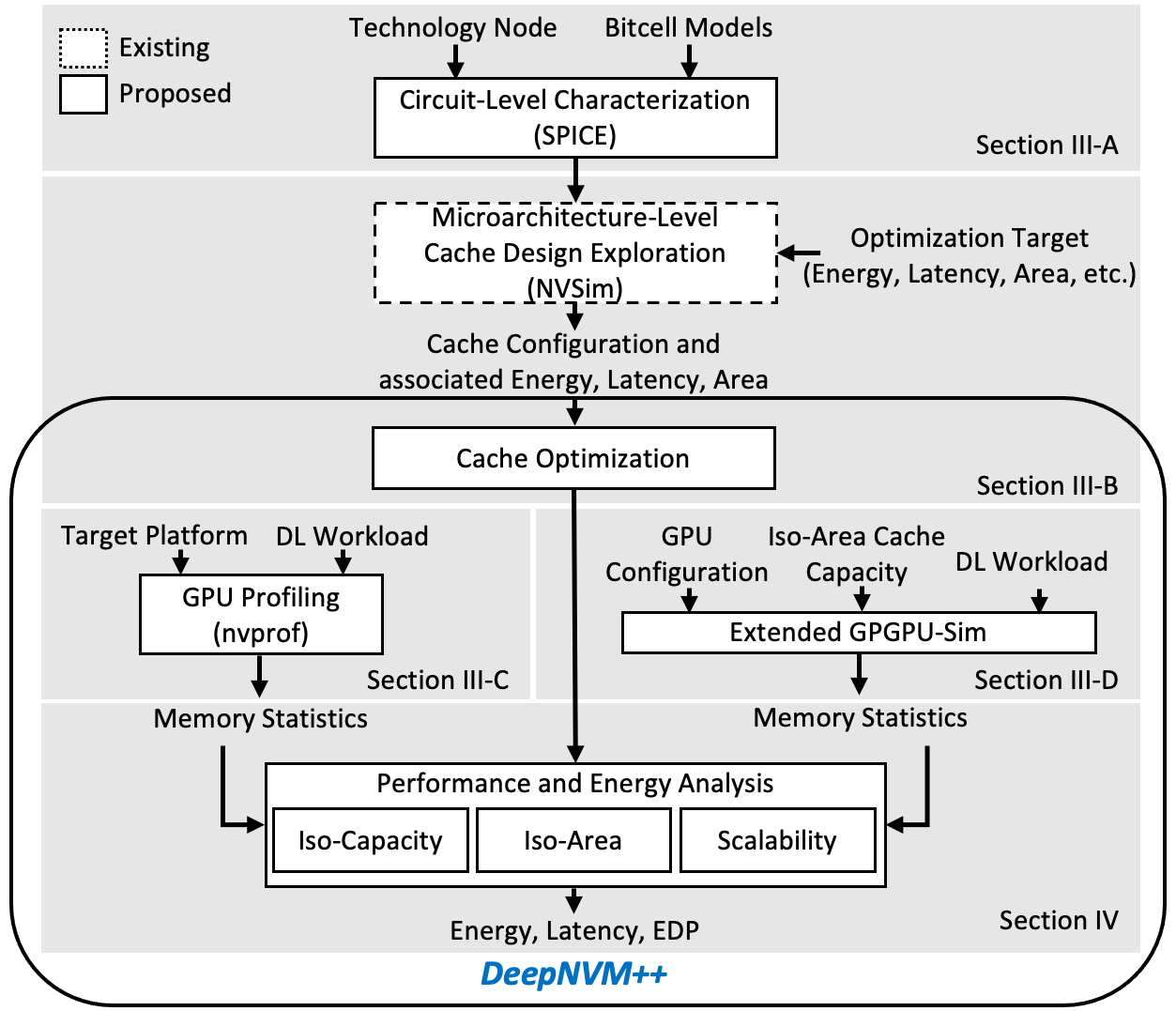}}
 \caption{Overview of the cross-layer analysis flow}
 \label{fig:flow}
\end{figure}

Prior work has proposed effective approaches to overcome the shortcomings of emerging NVM technologies such as using hybrid SRAM and NVM-based caches that utilize the complementary features of different memory technologies \cite{hybridreg,hybrid_cache_isca09,imani2016hybrid,beigi20163dhybrid}, relaxing non-volatility properties to reduce the high write latency and energy \cite{nvm_retention,stt_adaptable_retention,cache_revive_retention,sun2011retention}, and implementing cache replacement policies \cite{nvm_replacement,3dmram2009hybrid,imani2016hybridnarrow} for higher level caches such as L1 caches and register files. However, NVM technology appear to be a better choice for lower level caches such as L2 or L3 caches due to its long write latency and high cell density. Higher level L1 caches are latency-sensitive and optimized for performance, whereas last-level caches are capacity-sensitive and optimized for a high hit rate to reduce off-chip memory accesses. Therefore, NVM-based caches provide a better use case for replacing SRAM in last-level caches due to their high cell density when compared to SRAM-based caches. To this end, we evaluate power, performance, and area of NVM technology when used for last-level caches in GPU platforms.

While prior work has shown the potential of NVM technologies for generic applications to some extent, there is a need for a cross-layer analysis framework to explore the potential of NVM technologies in GPU platforms, particularly for DL workloads.
The most commonly used modeling tool for emerging NVM technologies is \textit{NVSim} \cite{nvsim}, a circuit-level model for performance, energy, and area estimation. However, \textit{NVSim} is not sufficient to perform a detailed cross-layer analysis for NVM technologies for DL workloads since it does not take architecture-level analysis and application-specific memory behavior into account. In this paper, we incorporate \textit{NVSim} with our cross-layer modeling and optimization flow including novel architecture-level iso-capacity and iso-area analysis flow to perform design space exploration for conventional SRAM and emerging NVM caches for DL workloads. 
This paper makes the following novel contributions:

\begin{enumerate}
    \item \textbf{Circuit-level bitcell characterization.} We perform detailed circuit-level characterization combining a commercial 16nm CMOS technology and prominent STT \cite{stt_model} and SOT \cite{sot_model} models from the literature to iterate through our framework in an end-to-end manner to demonstrate the flexibility of \textit{DeepNVM++} for future studies.
    \item \textbf{Microarchitecture-level cache design exploration.} We use \textit{NVSim} \cite{nvsim} to perform a fair comparison between SRAM, STT-MRAM, and SOT-MRAM by incorporating the circuit-level models developed in 1) using 16nm technology and choosing the best cache configuration for each of them.
    \item \textbf{Iso-capacity analysis.} To compare the efficacy of MRAM caches to conventional SRAM caches, we perform our novel iso-capacity analysis based on \textit{actual platform profiling} results for the memory behavior of various DNNs by using the \textit{Caffe} framework \cite{caffe} on a high-end NVIDIA 1080 Ti GPU (implemented in 16nm technology) for the ImageNet dataset \cite{imagenet}. 
    \item \textbf{Iso-area analysis.} Because of their different densities, we compare SRAM and NVM caches in an iso-area analysis to quantify the benefits of higher density of NVM technologies on DL workloads running on GPU platforms. Since existing platforms do not support resulting iso-area cache sizes, we extend the GPGPU-Sim~\cite{gpgpu-sim} simulator to run DL workloads and support larger cache capacities for STT-MRAM and SOT-MRAM. 
    \item \textbf{Scalability analysis.} Finally, we perform a thorough scalability analysis and compare SRAM, STT-MRAM, and SOT-MRAM in terms of power, performance, and area to project and gauge the efficacy of NVM and SRAM-based caches for DL workloads as cache capacity increases.
    \end{enumerate}

To the best of our knowledge, putting everything together, \textit{DeepNVM++} is the \textit{first comprehensive framework} for cross-layer characterization, modeling, and analysis of emerging NVM technologies for deep learning workloads running on GPU platforms. Our results show that in the iso-capacity case, STT-MRAM and SOT-MRAM achieve up to \textit{$3.8 \times$ and $4.7 \times$ energy-delay product reduction} and \textit{$2.4 \times$ and $2.8 \times$ area reduction} compared to SRAM baseline, respectively. In the iso-area case, STT-MRAM and SOT-MRAM achieve up to \textit{$2 \times$ and $2.3 \times$ energy-delay product reduction} and accommodate \textit{$2.3 \times$ and $3.3 \times$ cache capacity} compared to SRAM, respectively.

The rest of the paper is organized as follows. In Section III, we describe the details of our methodology from circuit to microarchitecture-level characterization, modeling, and analysis to obtain SRAM, STT-MRAM, and SOT-MRAM cache parameters. We also detail our iso-capacity and iso-area analysis methodology. In Section IV, we show experimental results demonstrating the efficiency of STT-MRAM and SOT-MRAM over the conventional SRAM for iso-capacity and iso-area cases. Furthermore, we perform a scalability analysis and show the PPA of SRAM, STT-MRAM, and SOT-MRAM. Next, we discuss the implications of the results shown in this paper in Section V. Finally, Section VI concludes the paper by summarizing the results.


\section{Methodology}
\begin{table}[t]
\small
\centering
\caption{{{{STT-MRAM and SOT-MRAM bitcell parameters after device level characterization}}}}
\label{table:powertable}
\scalebox{0.9}{
\begin{tabular}{|l|c|c|}
\hline
 & STT-MRAM & SOT-MRAM \\ \hline
Sense Latency (ps) & 650 & 650 \\ \hline
Sense Energy (pJ) & 0.076 & 0.020 \\ \hline
Write Latency (ps) & 8400 (set) / 7780 (reset) & 313 (set) / 243 (reset) \\ \hline
Write Energy (pJ) & 1.1 (set) / 2.2 (reset) & 0.08 (set) / 0.08 (reset) \\ \hline
Fin Counts & 4 (read/write) & 3 (write) + 1 (read) \\ \hline
Area (normalized) & 0.34* & 0.29* \\ \hline
\end{tabular}}
\begin{tablenotes}
\item[1] *: Area is normalized with respect to the foundry SRAM bitcell
\end{tablenotes}
\end{table}

In this section, we present our cross-layer analysis framework, as shown in Figure~\ref{fig:flow}. First, we show our detailed circuit-level characterization analysis using CMOS, STT, and SOT device models (Section III-A). After developing bitcell models, we present our microarchitecture-level cache design methodology to obtain cache area, latency, and energy results (Section III-B). Next, we describe our iso-capacity analysis flow in which we gather actual memory statistics through GPU profiling (Section III-C). Finally, we detail our iso-area analysis in which we extend GPGPU-Sim to run deep learning workloads and support larger cache capacities for STT-MRAM and SOT-MRAM (Section III-D).

\subsection{Circuit-level NVM Characterization}

A vast majority of work in the literature uses simple bitcell models \cite{mehdiNVM} to assess the PPA of corresponding cache designs. Because bitcells are the core components of the memory, the methodology to calculate the bitcell latency, energy, and area is crucial for accurate comparisons. To this end, we use a commercial 16nm bitcell design as a baseline as we model the STT and SOT bitcells. This technology node also matches the fabrication technology of the GPU platform that we use to gather actual memory statistics in Section III-C. 

The key bitcell parameters needed for cache modeling are read and write currents and latency values for high-to-low and low-to-high resistive transitions. These parameters can be optimized by tuning the size of the access transistors. While larger access transistors enable faster reads and writes, they increase the energy consumption and the bitcell layout size. 

For our simulations, we used perpendicular to the plane STT \cite{stt_model} and SOT \cite{sot_model} models and a commercial 16nm FinFET model that takes post-layout effects into account. To find the latency and energy parameters, we used parametrized SPICE netlists wherein the read/write pulse widths were modulated to the point of failure. Furthermore, we swept a range of fin counts for the access devices to find the optimal balance between the latency, energy, and area. To calculate the bitcell area for the 16nm layout design rules, we used the bitcell area formulations provided in prior work \cite{kaushik_roy_sot_area}.

We summarize the obtained bitcell parameters in Table~\ref{table:powertable}. The sensing delay is measured from wordline activation to the point where the bitline voltage difference reaches 25mV. The sense energy is the integration of the power consumed over the sensing time window. For both magnetic flavors, the sense delay is similar; however, SOT-MRAM is more energy-efficient in terms of read operation owing to the separation of the read/write terminals. The write latency in this context refers to the time between the arrival of the write enable signal to the access transistor and a complete magnetization change for the MTJ. The write latencies for STT and SOT bitcells are significantly different, as expected. This difference can be seen in the energy values as well. A bigger access device is needed in order to enable a larger current flow to the STT bitcell. The isolation of the read and the write terminals in the SOT bitcell allows for a smaller write access device. The area values are normalized by the foundry bitcell area. We highlight the significant area difference and demonstrate its impact on the cache characteristics in Section III-B. We use these bitcell parameters for energy-delay-area product (EDAP) optimized cache design exploration as discussed in the next section.

\subsection{Microarchitecture-level Cache Design Exploration}

\begin{algorithm}[h]
 \SetAlgoLined
 $mem \in \mathcal M = \{SRAM,STT,SOT\}$\; 
 $cap \in \mathcal C = \{1,2,4,8,16,32\}$\;
 $opt \in \mathcal O = \{Read_{Latency},Write_{Latency},Read_{Energy},$\\ $...Write_{Energy},Read_{EDP},Write_{EDP},Area,Leakage\}$\;
  \vspace{-4mm}
 $acc \in \mathcal A = \{Normal,Fast,Sequential\}$\;

 \For{$\textbf{each} \hspace{5pt}  mem \in \mathcal M $}{
    \For{$\textbf{each} \hspace{5pt} cap \in \mathcal C $}{
        $ Q^{'} \leftarrow \infty $\; 
        \For{$\textbf{each} \hspace{5pt} opt \in \mathcal O $}{
            \For{$\textbf{each} \hspace{5pt} acc \in \mathcal A $}{
            $Q \leftarrow calculate(EDAP)$\; 
            \If{$Q < Q^{'}$}{
                $ Q^{'} \leftarrow Q $\; 
                }
            }
        }
    $TunedConfig.append(argv(Q))$\;
    }
 }
\textbf{return} $TunedConfig$\;
\caption{EDAP-Optimal Cache Tuning Algorithm}
\end{algorithm}

In order to demonstrate the impact of using STT and SOT bitcells in L2 caches, we use \textit{NVSim} \cite{nvsim}, a circuit-level analysis framework that delivers energy, latency, and area results. After developing \textit{NVSim}-compatible bitcell models as described in Section III-A, we analyzed a range of cache capacities (1MB to 32MB) for all possible configurations and cache access types to demonstrate the potential of STT-MRAM and SOT-MRAM as the cache capacity tends to grow. Such a scalability study will help in determining the benefits of switching from conventional SRAM to NVM-based caches in future GPU platforms as depicted by the trend in Figure~\ref{fig:trend}.

Algorithm 1 depicts the EDAP-optimal cache tuning algorithm. Based on the optimization target used in \textit{NVSim}, the cache PPA values vary substantially. Therefore, we independently choose the best configuration for each type of memory technology in terms of EDAP metric to perform a fair comparison that encompasses all and not just one of the design constraint dimensions. 

As described in Section III-A, we use a commercial 16nm bitcell design. To ensure a correct analysis, we modified the internal technology file of \textit{NVSim} to the corresponding 16nm technology parameters. Next, we compare SRAM, STT-MRAM, and SOT-MRAM for various cache capacities in terms of area, latency, and energy results. Based on these, we determine the EDAP for the cache (as denoted by \textit{calculate(EDAP)} in Algorithm 1). 

\begin{table}[t]
\small
\centering
\caption{{Latency, energy, and area results for SRAM, STT-MRAM, and SOT-MRAM caches for iso-capacity and iso-area}}
\label{table:xtable}
\scalebox{0.75}{
\begin{tabular}{|c|c|c|c|c|c|}
\hline
\multirow{2}{*}{} & \multirow{2}{*}{SRAM} & \multicolumn{2}{c|}{STT-MRAM} & \multicolumn{2}{c|}{SOT-MRAM} \\ \cline{3-6} 
 &  & Iso-Capacity & Iso-Area & Iso-Capacity & Iso-Area \\ \hline
Capacity (MB) & 3 & 3 & 7 & 3 & 10 \\ \hline
Read Latency (ns) & 2.91 & 2.98 & 4.58 & 3.71 & 6.69 \\ \hline
Write Latency (ns) & 1.53 & 9.31 & 10.06 & 1.38 & 2.47 \\ \hline
Read Energy (nJ) & 0.35 & 0.81 & 0.93 & 0.49 & 0.51 \\ \hline
Write Energy (nJ) & 0.32 & 0.31 & 0.43 & 0.22 & 0.40 \\ \hline
Leakage Power (mW) & 6442 & 748 & 1706 & 527 & 1434 \\ \hline
Area (mm\textsuperscript{2}) & 5.53 & 2.34 & 5.12 & 1.95 & 5.64 \\ \hline
\end{tabular}}
\end{table}

\begin{figure*}[h]
  \centering
 \subfloat{\includegraphics[width=0.49\textwidth]{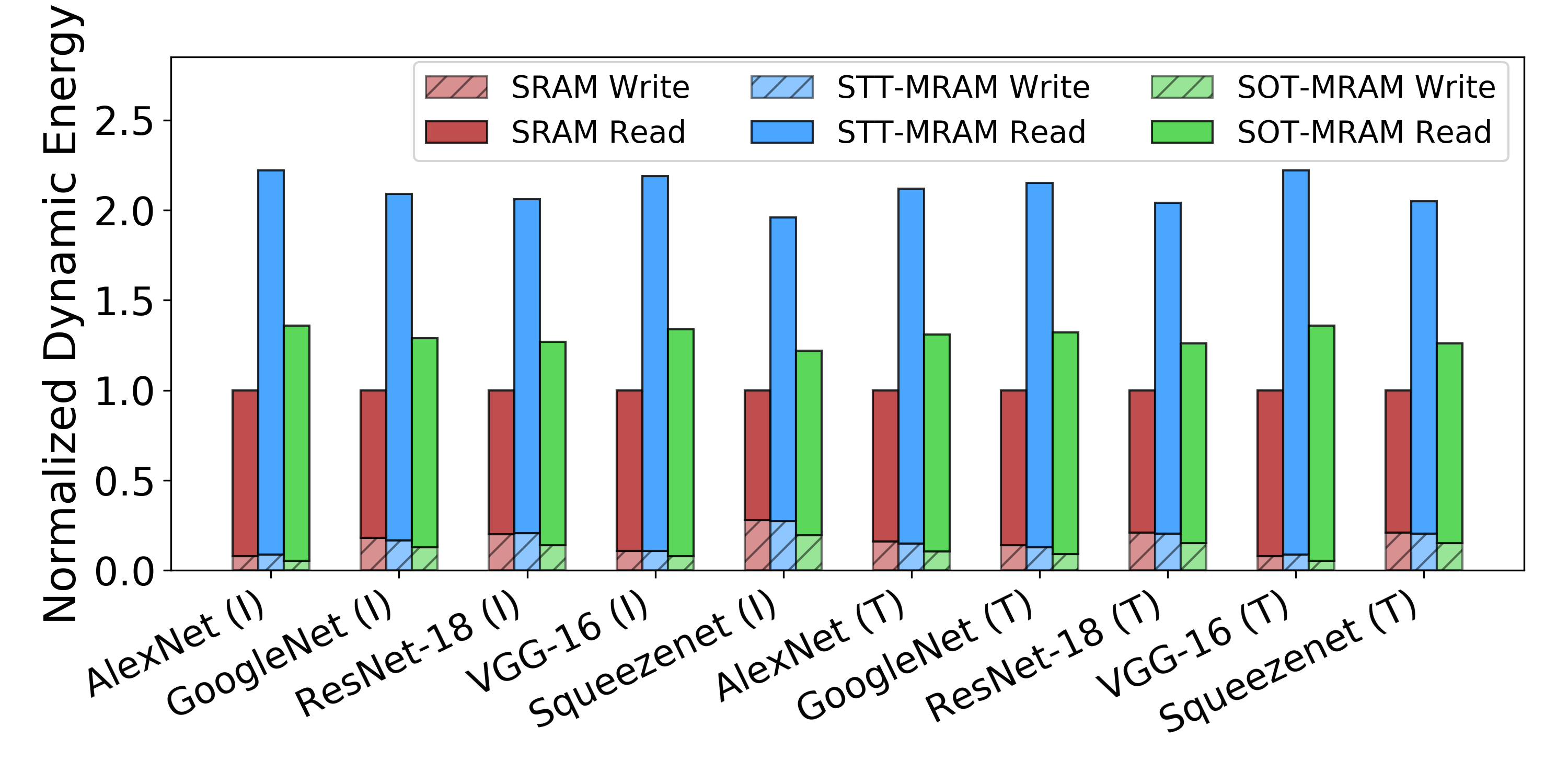}}
  \subfloat{\includegraphics[width=0.49\textwidth]{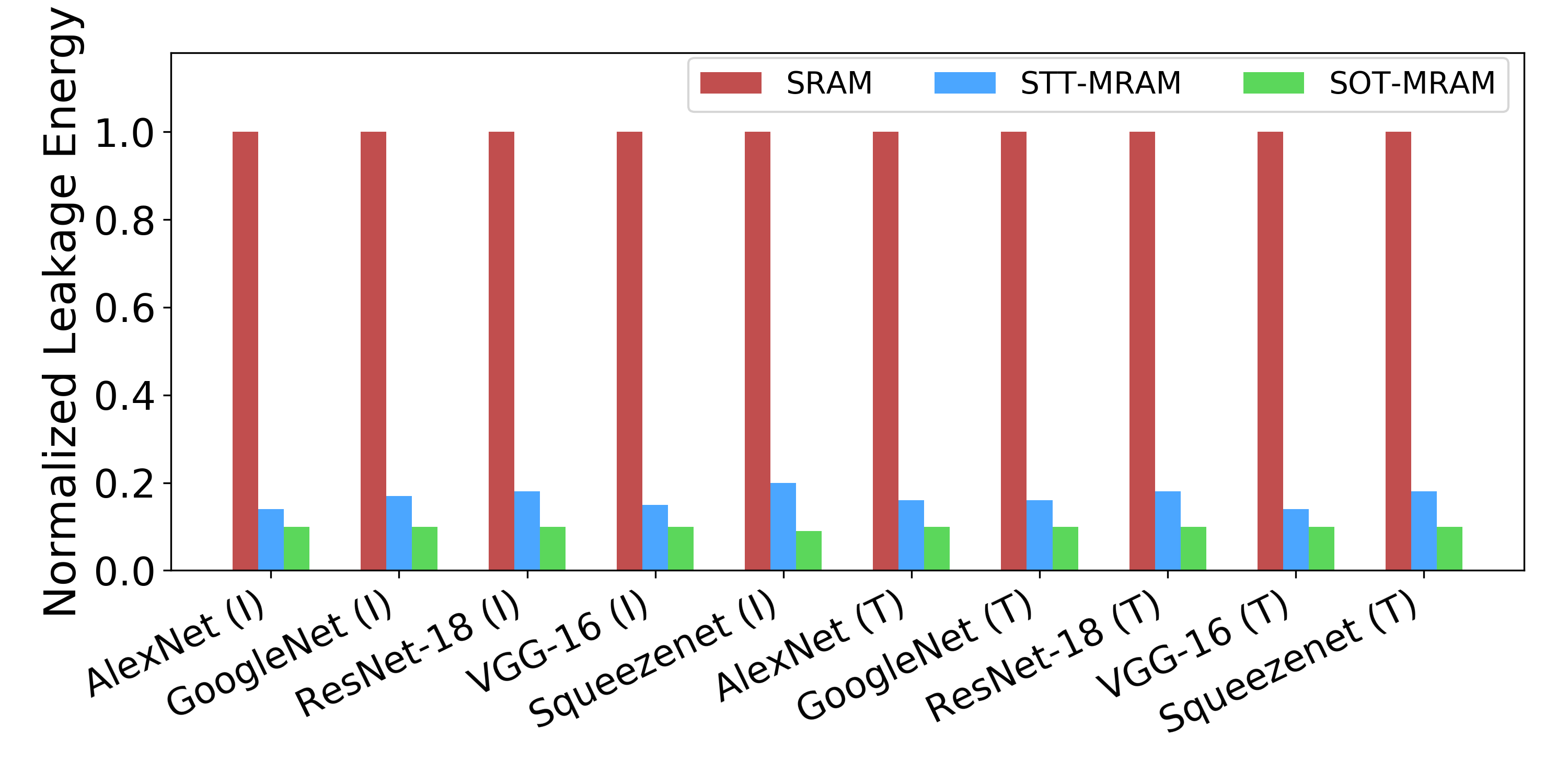}}
 \caption{Dynamic energy (left c energy (right chart) (lower is better) normalized with respect to SRAM by using NVMs with iso-capacity (3MB) for inference (I) and training (T) stages}
 \label{fig:norm_nets_energy}
\end{figure*}

\begin{figure*}[h]
    \centering
      \subfloat{\includegraphics[width=0.49\textwidth]{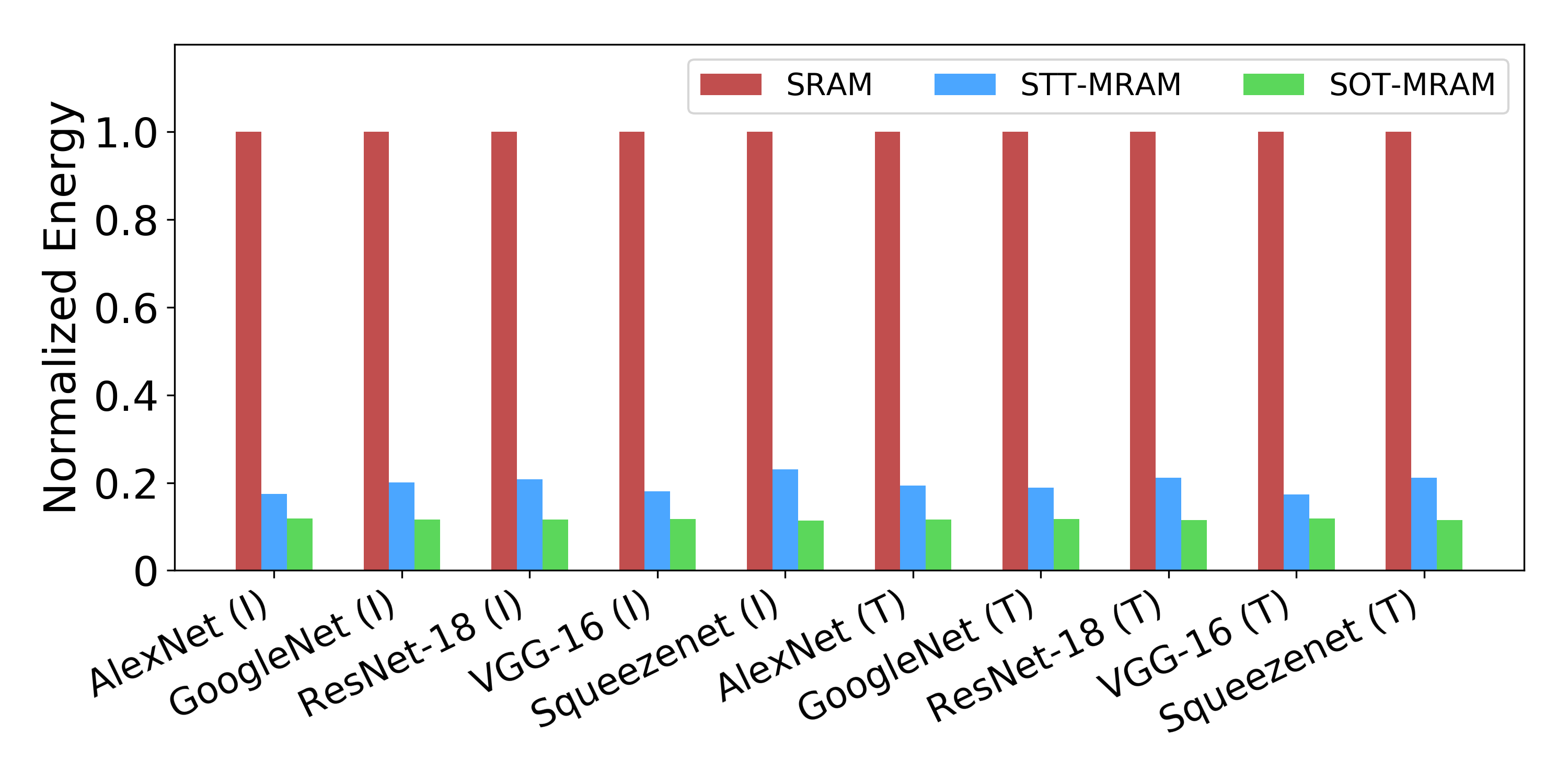}}
  \subfloat{\includegraphics[width=0.49\textwidth]{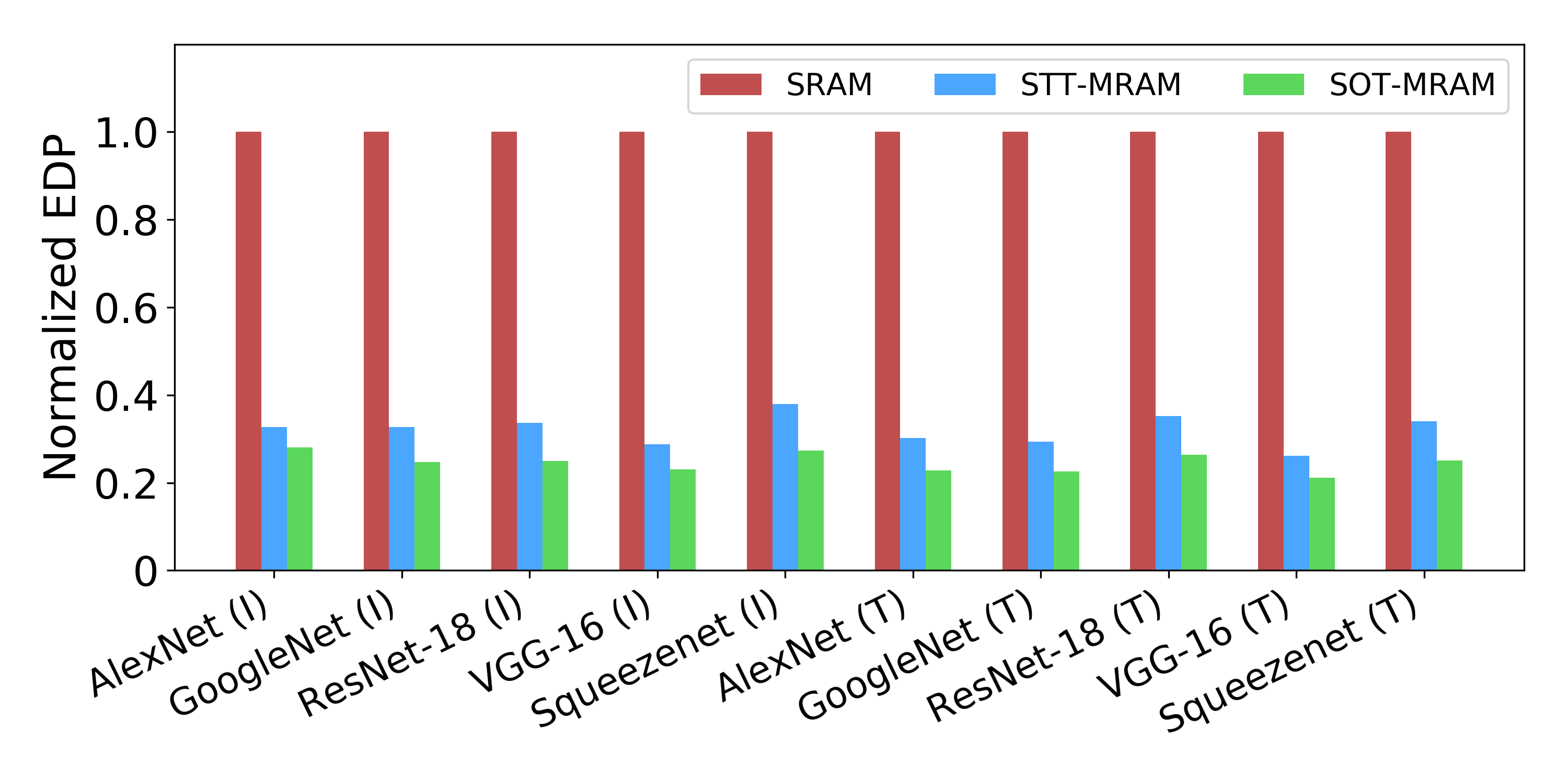}}
    \caption{Iso-capacity (3MB) energy and energy-delay product for NVM-based caches (lower is better) normalized with respect to SRAM-based caches for inference (I) and training (T) stages. DRAM energy and latency are also included in EDP results.}
    \label{fig:norm_energy_leakage_dynamic}
\end{figure*}

Table~\ref{table:xtable} shows the latency, energy, and area results that correspond to the cache capacity of 1080 Ti GPU (3MB) and to the larger MRAM caches that fit into the same area of SRAM baseline. We convert read and write latencies to clock cycles based on 1080 Ti GPU's clock frequency for our calculations. For STT-MRAM and SOT-MRAM, we show parameters for both iso-capacity and iso-area when compared to SRAM. We use these parameters to evaluate the workload dependent impact of memory choices using DL workloads with diverse structures and multiply-accumulate operations (MACs) configurations. 

The energy and latency benefits of STT-MRAM and SOT-MRAM depend on the data characteristics of a given workload. To account for differences in the data-related read/write characteristics, we used a simple model where we multiply the number of read and write transactions by the corresponding latency and energy values for those operations. 

\textbf{Implications in architecture-level analysis.} To gauge the benefits of using MRAM technology, we consider two scenarios: (i) First, one could replace the SRAM cache in a GPU with the same capacity MRAM with a smaller area. (ii) Alternatively, by using the same area dedicated to the cache, one can increase the on-chip cache capacity, thereby reducing costly DRAM traffic. We analyze and discuss both approaches through platform profiling results for iso-capacity scenario and a set of architecture-level simulations for iso-area scenario.

\subsection{Architecture-level Iso-Capacity Analysis}

\begin{table}[h]
\centering
\caption{Configurations for DNNs under consideration}
\label{table:dnn}
\scalebox{0.85}{
\begin{tabular}{|c|c|c|c|c|c|}
\hline
 & \textbf{AlexNet} & \textbf{GoogLeNet} & \textbf{VGG-16} & \textbf{ResNet-18} & \textbf{SqueezeNet}\\ \hline
Top-5 error & 16.4 & 6.7 & 7.3 & 10.71 & 16.4 \\ \hline
CONV Layers & 5 & 57 & 13 & 17 & 26 \\ \hline
FC Layers & 3 & 1 & 3 & 1 & 0 \\ \hline
Total Weights & 61M & 7M & 138M & 11.8M & 1.2M \\ \hline
Total MACs & 724M & 1.43G & 15.5G & 2G &  837M \\ \hline
\end{tabular}
}
\end{table}

As the platform target to demonstrate our work, we use a high-end 1080 Ti GPU which is fabricated in a commercial 16nm technology node which also matches our bitcell and cache models. 
We use the \textit{Caffe} \cite{caffe} framework to run various DNNs such as AlexNet \cite{alexnet}, GoogLeNet \cite{googlenet}, VGG-16 \cite{vgg16}, ResNet-18 \cite{resnet50}, and SqueezeNet \cite{squeezenet} for the ImageNet \cite{imagenet} dataset as shown in Table III. Our analysis is generalizable to other types of neural network architectures since we cover a wide range of DNN configurations with various workload characteristics. We use the NVIDIA profiler \cite{nvprof} to obtain the device memory and L2 cache read and write transactions to better understand both on-chip and off-chip memory behavior of DNN workloads.

\subsection{Architecture-level Iso-Area Analysis}

Since the iso-area larger capacities enabled by higher density NVM implementations do not exist in existing platforms, we use \textit{GPGPU-Sim}~\cite{gpgpu-sim} to explore power and performance implications of having these larger L2 caches in GPU architectures for DNN workloads. For comparison, we model the high-end GTX 1080 Ti GPU. The configurations for 1080 Ti GPU are shown in Table~\ref{table:gpgpusimconfig}. We extend the \textit{GPGPU-Sim} simulator to support the cache capacity of GTX 1080 Ti GPU. This GPU is built using a commercial 16nm technology node which matches our bitcell and cache models. In particular, for \textit{GPGPU-Sim} compatibility, we set L2 cache capacity to 3MB. We use this capacity for our analysis in the rest of the paper. We measure the number of DRAM transactions to quantify and better understand the relationship between larger L2 caches and the overall system power and performance. As a DNN benchmark, we use AlexNet \cite{alexnet} with the ImageNet \cite{imagenet} dataset which is provided by the \textit{DarkNet} \cite{darknet} framework. We extend \textit{DarkNet} source code to enable deep learning workloads on \textit{GPGPU-Sim}.\footnote{We plan to open source our framework after publication to enable future research on GPU architectures for deep learning workloads.}

\begin{table}[t]
\centering
\caption{GPGPU-Sim Configurations}
\label{table:gpgpusimconfig}
\scalebox{0.9}{
\begin{tabular}{|c|c|}
\hline
 & GTX 1080 Ti \\ \hline
Number of Cores & 28 \\ \hline
Number of Threads/Core & 2048 \\ \hline
Number of Registers/Core & 65536 \\ \hline
\multirow{2}{*}{L1 Data Cache} & 48 KB, 128 B line, \\
 & 6-way LRU \\ \hline
\multirow{2}{*}{L2 Data Cache} & 128 KB/channel, 128 B \\
 & line, 16-way LRU \\ \hline
\multirow{2}{*}{Instruction Cache} & 8 KB, 128 B line, \\
 & 16-way LRU \\ \hline
Number of & \multirow{2}{*}{4} \\
Schedulers / Core &  \\ \hline
Frequency (MHz): & \multirow{2}{*}{1481, 2962,} \\
Core, Interconnect, & \multirow{2}{*}{1481, 2750} \\ 
L2, Memory &  \\ \hline
\end{tabular}}
\end{table}

\begin{figure*}[t]
  \centering
 \subfloat{\includegraphics[width=0.49\textwidth]{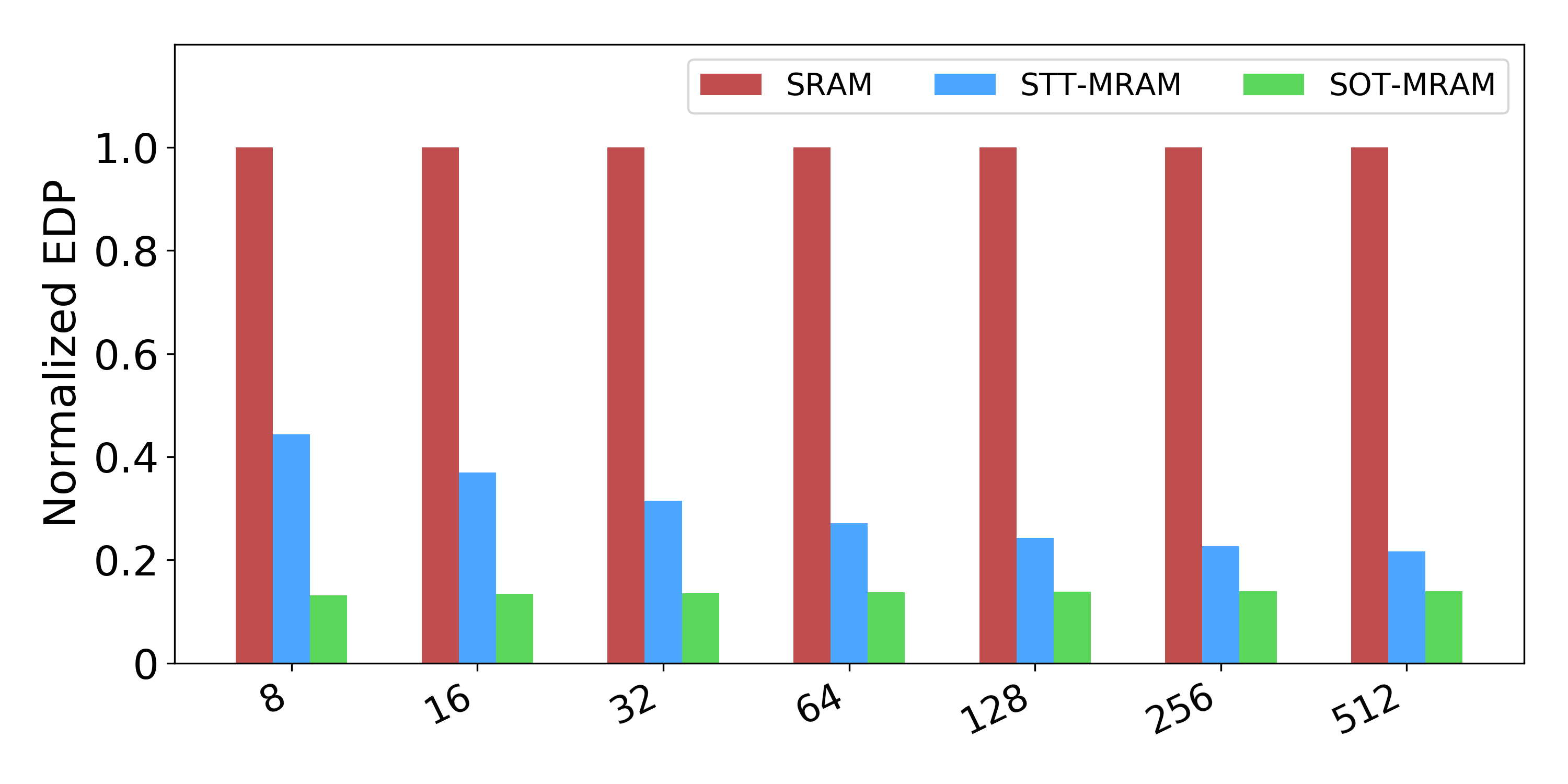}}
  \subfloat{\includegraphics[width=0.49\textwidth]{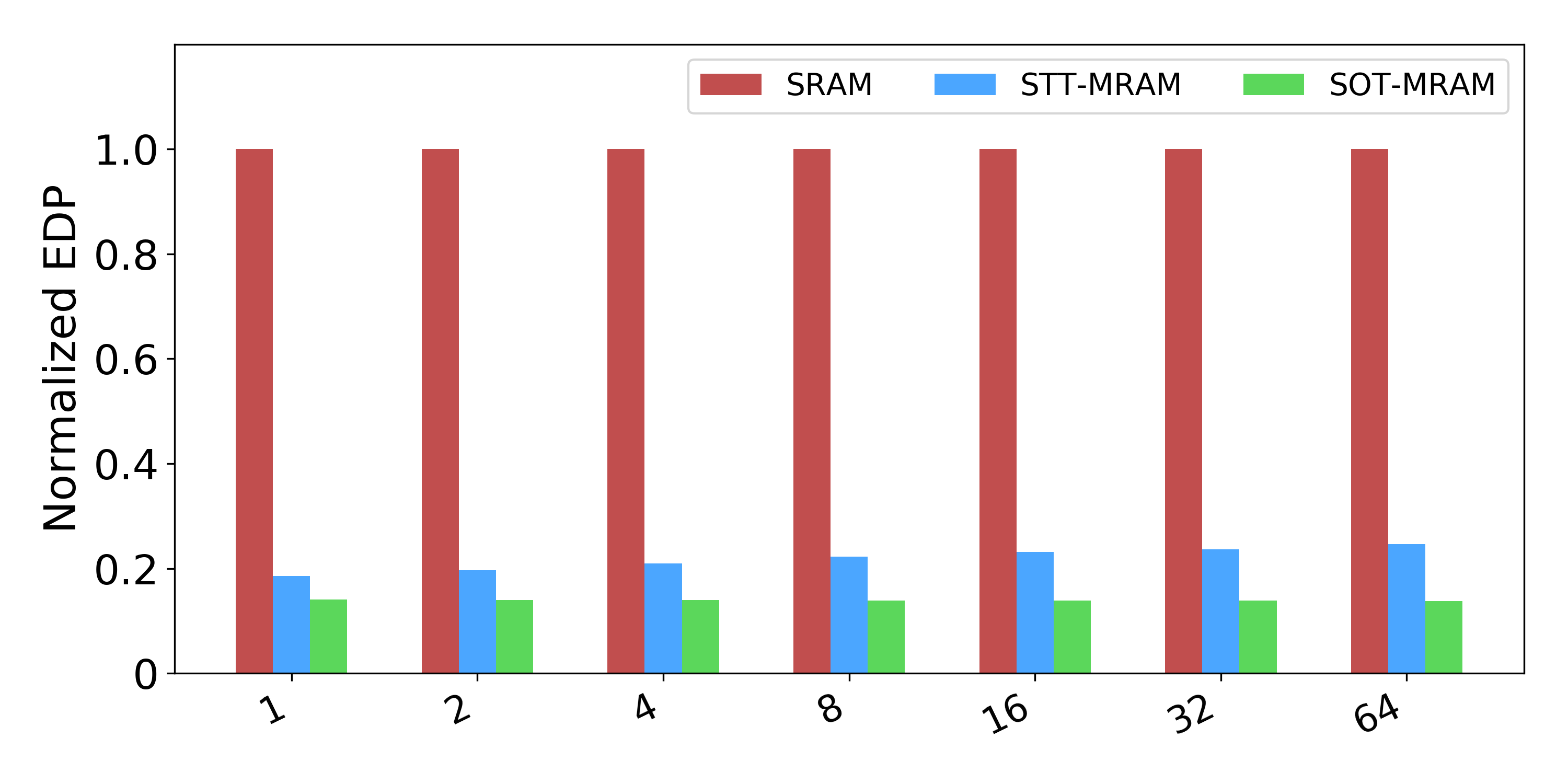}}
 \caption{Impact of batch size on energy-delay product (lower is better) normalized with respect to SRAM by using NVMs with iso-capacity (3MB) for AlexNet for training (left chart) and inference (right chart)}\label{fig:batchsize}
\end{figure*}

\section{Results}
We analyze STT-MRAM and SOT-MRAM in terms of energy, performance, and area results by using GPU profiling results for both iso-capacity and iso-area cases in Section IV-A and Section IV-B, respectively. In Section IV-B, we use iso-area cache parameters as shown in Table~\ref{table:xtable} and we use \textit{GPGPU-Sim} to quantify the DRAM access reduction in the iso-area case at larger cache capacities. We include DRAM accesses in our performance and energy calculations for iso-area case. In Section IV-C, we perform a scalability analysis to project the implications of the current GPU trend shown in Figure~\ref{fig:trend} on performance and energy results.

\subsection{Performance and Energy Results for Iso-Capacity}

By combining the actual technology-dependent latency and energy metrics from Table~\ref{table:xtable}, we can perform a performance and energy analysis for replacing conventional SRAM caches with MRAM caches. We choose batch size 4 for inference and 64 for training for our workloads as it is typically used in related work \cite{eyeriss_ssc}.

Figure~\ref{fig:norm_nets_energy} shows normalized dynamic energy and leakage energy breakdown results for 1080 Ti GPU based on actual platform memory statistics and our MRAM cache models at the same cache capacity. We use our cache parameters and profiling results to calculate results for various DNNs for both inference and training workloads.

In Figure~\ref{fig:norm_nets_energy}, we observe that STT-MRAM has $2.1 \times$ dynamic energy whereas SOT-MRAM has $1.3 \times$ dynamic energy on average when compared to SRAM baseline. Furthermore, our results show that 83\% of the total dynamic energy of SRAM comes from read operations whereas write operations only make for 17\% of all transactions on average across all workloads. Our profiling results also support these findings as read operations dominate write operations in DL workloads.

On the other hand, Figure~\ref{fig:norm_nets_energy} also shows that STT-MRAM and SOT-MRAM provide $5.9 \times$ and $10 \times$ lower leakage energy on average when compared to SRAM, respectively. Based on this result, Figure~\ref{fig:norm_energy_leakage_dynamic} shows significant total normalized energy reduction of STT-MRAM and SOT-MRAM when compared to SRAM given that leakage energy dominates the total energy. In more detail, STT-MRAM and SOT-MRAM achieve \textit{$5.1 \times$ and $8.6 \times$ energy reduction} on average across all workloads compared to SRAM baseline, respectively, due to their significantly low leakage energy. Moreover, Figure~\ref{fig:norm_energy_leakage_dynamic} shows that STT-MRAM and SOT-MRAM provide up to \textit{$3.8 \times$ and $4.7 \times$ EDP reduction and $2.4 \times$ and $2.8 \times$ area reduction}, respectively.

\textbf{The impact of batch size on EDP.} We perform this study to better understand the relationship between batch size and its implications for performance and energy results of SRAM, STT-MRAM, and SOT-MRAM. Figure~\ref{fig:batchsize} shows the impact of batch size on EDP results for AlexNet during training and inference stages based on 1080 Ti memory profiling statistics. We show that batch size significantly affects the improvement of STT-MRAM and SOT-MRAM for training. For training, STT-MRAM provides $2.3 \times$ to $4.6 \times$ EDP reduction as batch size increases. On the other hand, SOT-MRAM provides $7.2 \times$ to $7.6 \times$ EDP reduction when compared to SRAM baseline. For inference, STT-MRAM and SOT-MRAM achieve $4.1 \times$ to $5.4 \times$ and $7.1 \times$ to $7.3 \times$ EDP reduction, respectively. These results also confirm the different workload characteristics of training and inference. STT-MRAM provides higher EDP reduction for training workloads as batch size increases. On the other hand, SOT-MRAM follows the same pattern for inference workloads due to their different access characteristics as shown in Table~\ref{table:xtable}. We observe that training workloads become more read dominant whereas inference workloads have lower read/write ratio as batch size increases.

\subsection{Performance and Energy Results for Iso-Area}

\begin{figure}[h]
  \centering
  \subfloat{\includegraphics[width=0.47\textwidth]{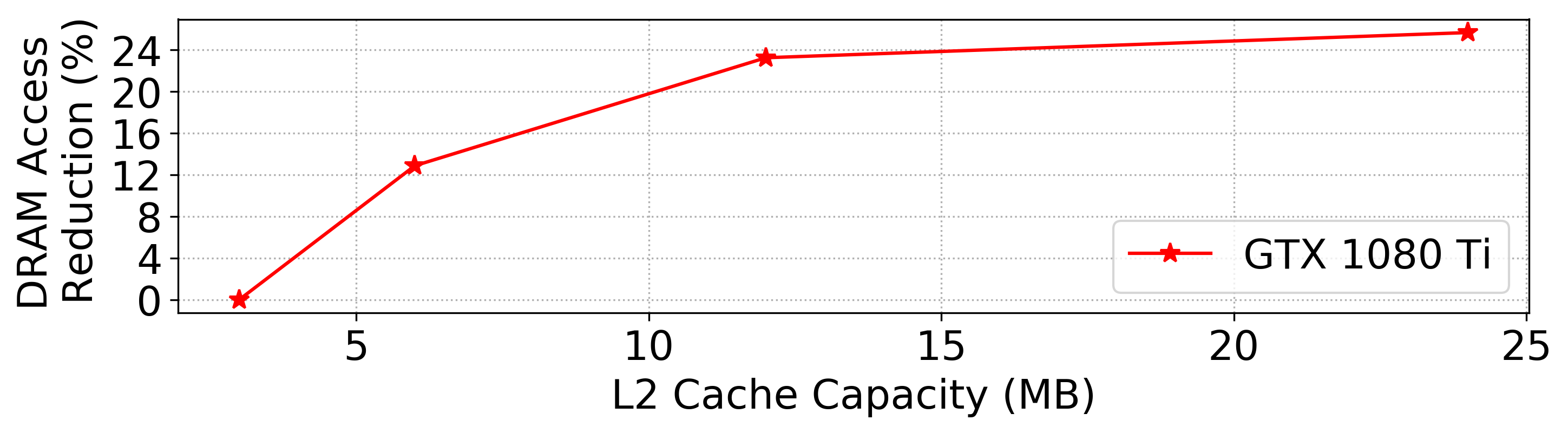}\label{fig:dram}}
 \caption{Simulation results for the reduction in the total number of DRAM accesses in percentage (\%)}
 \label{fig:gpgpusim}
\end{figure}

\begin{figure*}[!tbp]
  \centering
 \subfloat{\includegraphics[width=0.49\textwidth]{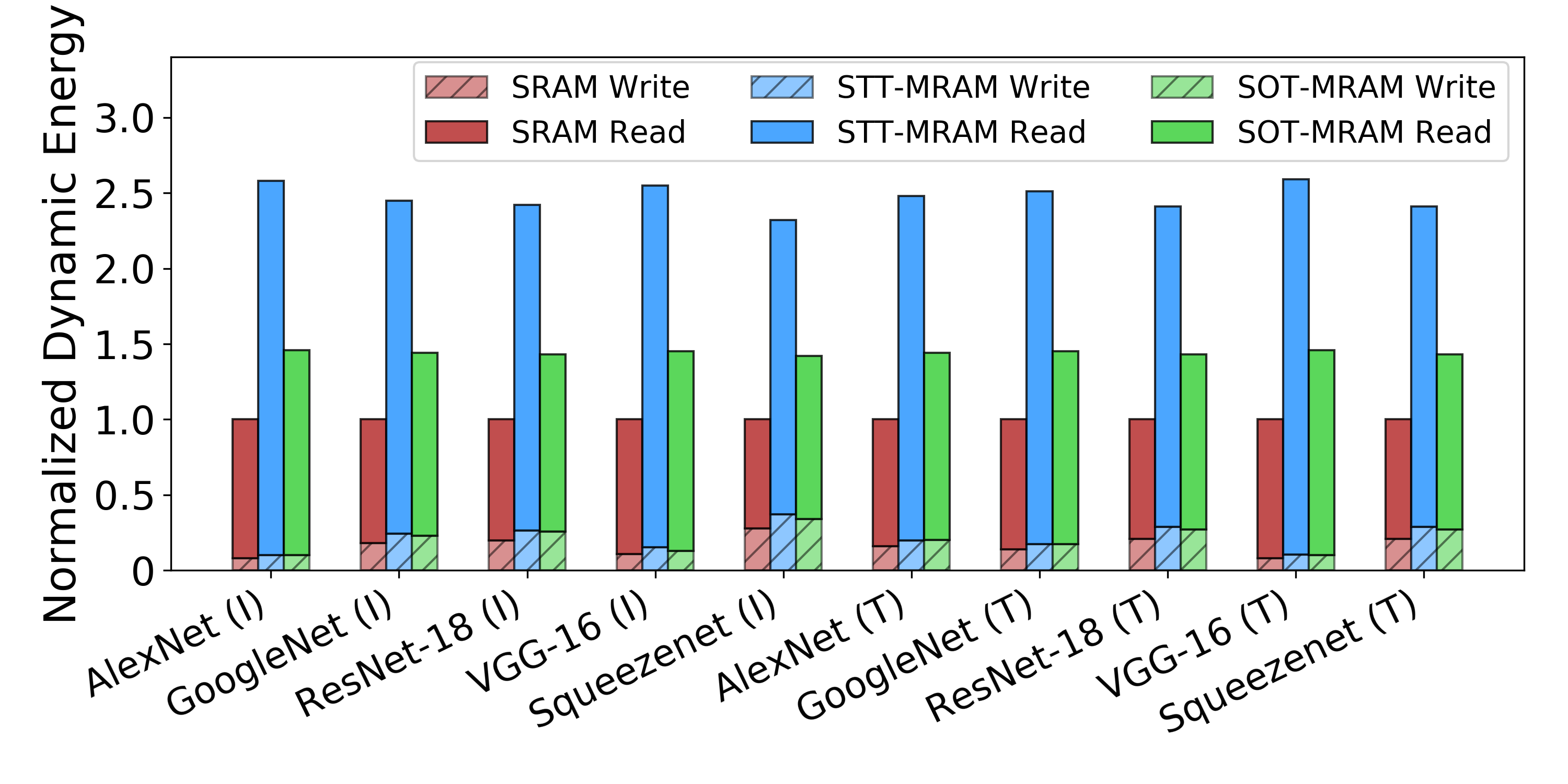}}
  \subfloat{\includegraphics[width=0.49\textwidth]{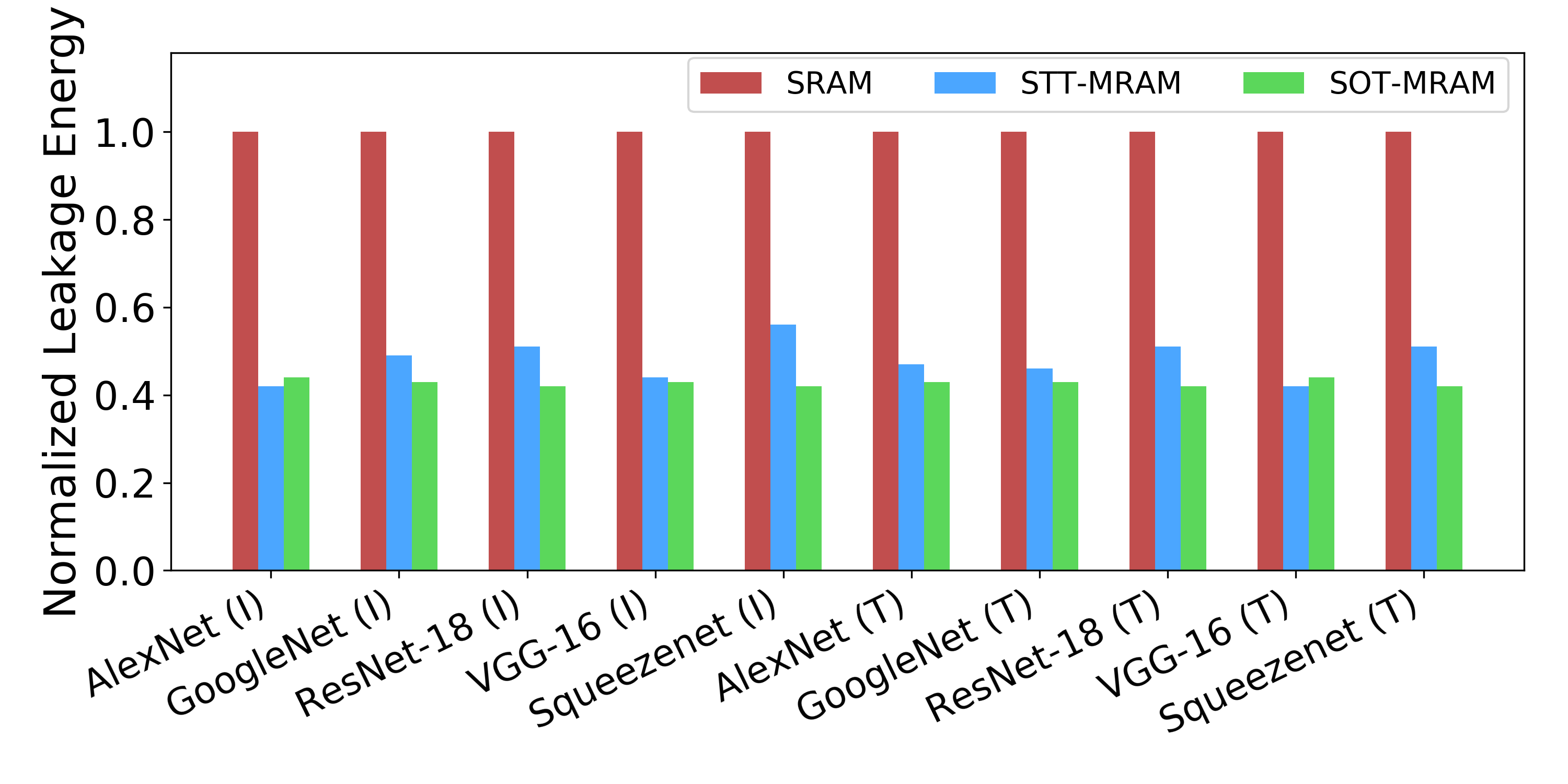}}
 \caption{Dynamic energy (left chart) and leakage energy (right chart) (lower is better) normalized with respect to SRAM by using STT-MRAM (7MB) and SOT-MRAM (10MB) with iso-area for inference (I) and training (T) stages}
 \label{fig:iso_area_norm_nets_energy}
\end{figure*}

\begin{figure*}[h]
  \centering
  \subfloat{\includegraphics[width=0.49\textwidth]{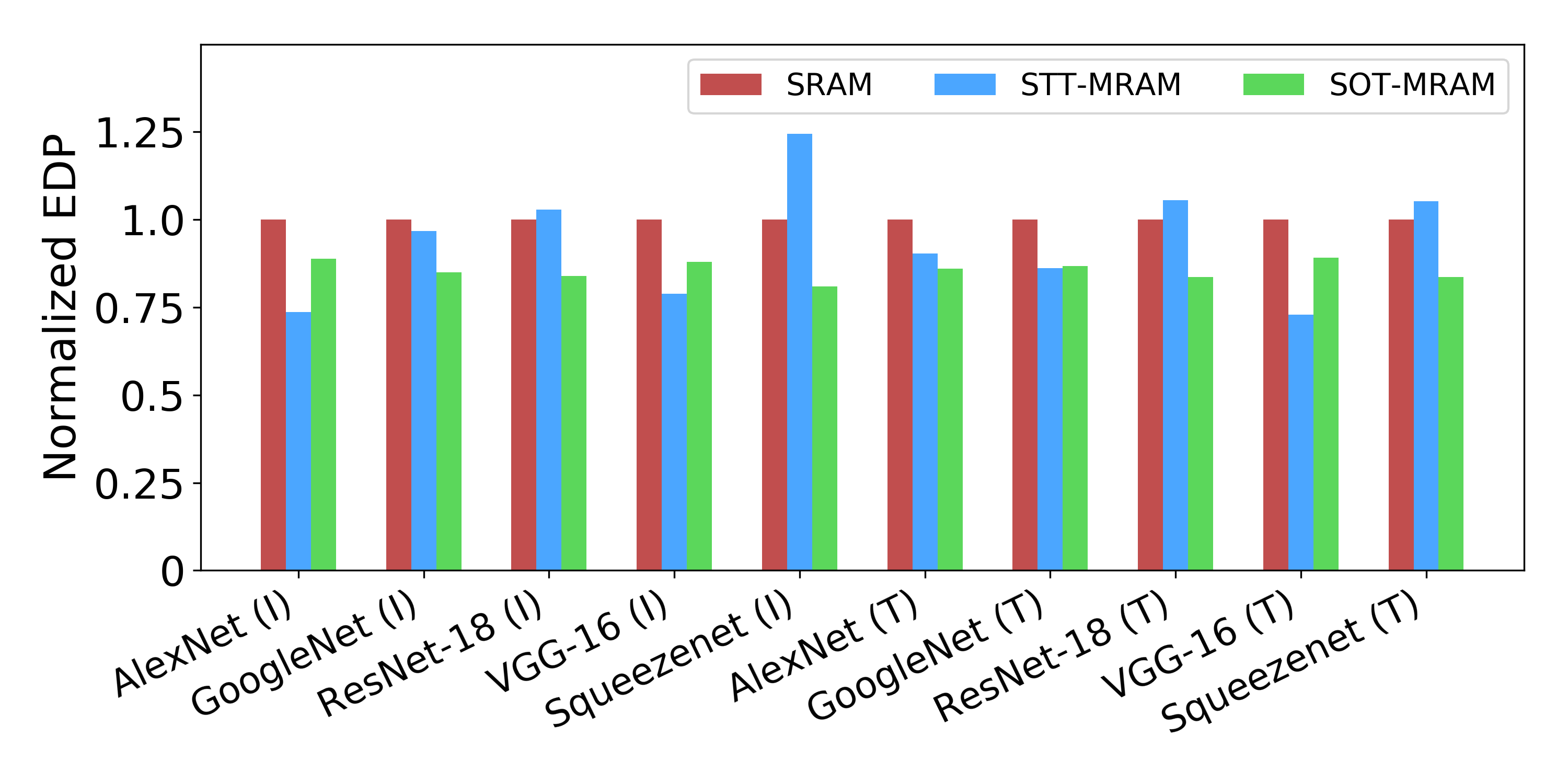}}
  \subfloat{\includegraphics[width=0.49\textwidth]{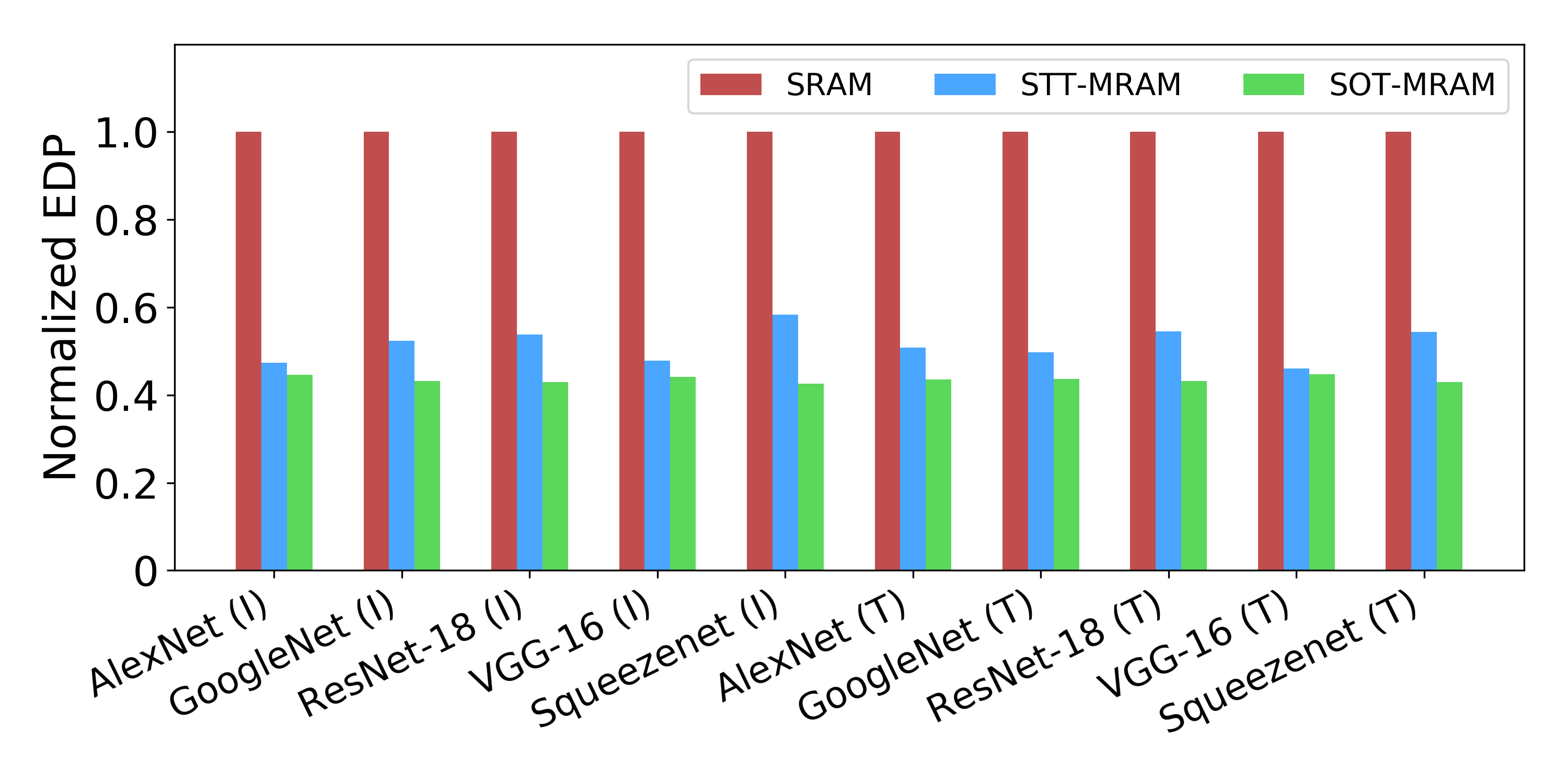}}
 \caption{Iso-area energy-delay product results for STT-MRAM (7MB) and SOT-MRAM (10MB) (lower is better) normalized with respect to SRAM-based caches for inference (I) and training (T) stages without (left chart) and with (right chart) DRAM energy and latency.}
 \label{fig:norm_edp_area}
\end{figure*}

As in the iso-capacity study, for iso-area analysis we use a batch size 4 for inference and 64 for training. Figure~\ref{fig:gpgpusim} shows the reduction in the total number of DRAM accesses as L2 cache capacity increases. We use \textit{GPGPU-Sim} and start with the baseline configuration which is 3MB for GTX 1080 Ti and double its cache capacity up to 24MB to quantify the percentage of DRAM access reduction for STT-MRAM and SOT-MRAM at larger cache capacities. Figure~\ref{fig:gpgpusim} shows that replacing SRAM with STT-MRAM and SOT-MRAM equivalents that fit into the same area significantly reduces the total number of DRAM transactions by 14.6\% and 19.8\%, respectively for 1080 Ti GPU. 

Figure~\ref{fig:iso_area_norm_nets_energy} shows normalized dynamic energy and leakage energy breakdown results for 1080 Ti GPU based on actual platform memory statistics and our MRAM cache models at the same area. We use our iso-area cache parameters in which STT-MRAM (7MB) and SOT-MRAM (10MB) have larger cache capacities for the same area budget with SRAM. We use these cache parameters and profiling results to calculate results for various DNNs for both inference and training workloads. 

In Figure~\ref{fig:iso_area_norm_nets_energy}, we observe that STT-MRAM has $2.5 \times$ dynamic energy whereas SOT-MRAM has $1.4 \times$ dynamic energy on average when compared to SRAM baseline. On the other hand, Figure~\ref{fig:iso_area_norm_nets_energy} also shows that STT-MRAM and SOT-MRAM provide $2.1 \times$ and $2.3 \times$ lower leakage energy on average when compared to SRAM, respectively. 
Based on this result, STT-MRAM and SOT-MRAM achieve \textit{$2 \times$ and $2.3 \times$ lower energy} when compared to SRAM.

Furthermore, Figure~\ref{fig:norm_edp_area} shows that STT-MRAM and SOT-MRAM provide \textit{$1.1 \times$ and $1.2 \times$ EDP reduction and $2.3 \times$ and $3.3 \times$ larger cache capacity} on average across all workloads when compared to SRAM and off-chip DRAM accesses are not included in the calculations, respectively. When DRAM accesses are included in determining EDP, as shown in Figure~\ref{fig:norm_edp_area}, STT-MRAM and SOT-MRAM provide \textit{$2 \times$ and $2.3 \times$ EDP reduction} on average across all workloads when compared to SRAM, respectively. 

We show that although the cache latency and energy results for STT-MRAM and SOT-MRAM do not outperform SRAM results at larger cache capacities as shown in Table~\ref{table:xtable}, they do outperform SRAM when costly off-chip DRAM accesses are also considered in EDP calculations. To this end, Chen \textit{et al}.~\cite{eyeriss} showed that the normalized energy cost of a global buffer access relative to a MAC operation is $6 \times$, whereas a DRAM access is $200 \times$ for a machine learning hardware accelerator. By the same token, the higher cell density of NVM can be exploited to shift the memory traffic from DRAM to L2 cache to further improve power and performance of the overall system. This approach can dramatically reduce the total number of costly DRAM accesses and reduce data movement, which is a daunting impediment for achieving energy-efficient machine learning hardware \cite{eyeriss_ssc,eyeriss,tetris,boroumand2018movement,donato2018onchipnvm}. 

\subsection{Scalability Analysis}

\begin{figure*}[!tbp]
  \centering
 \subfloat{\includegraphics[width=0.32\textwidth]{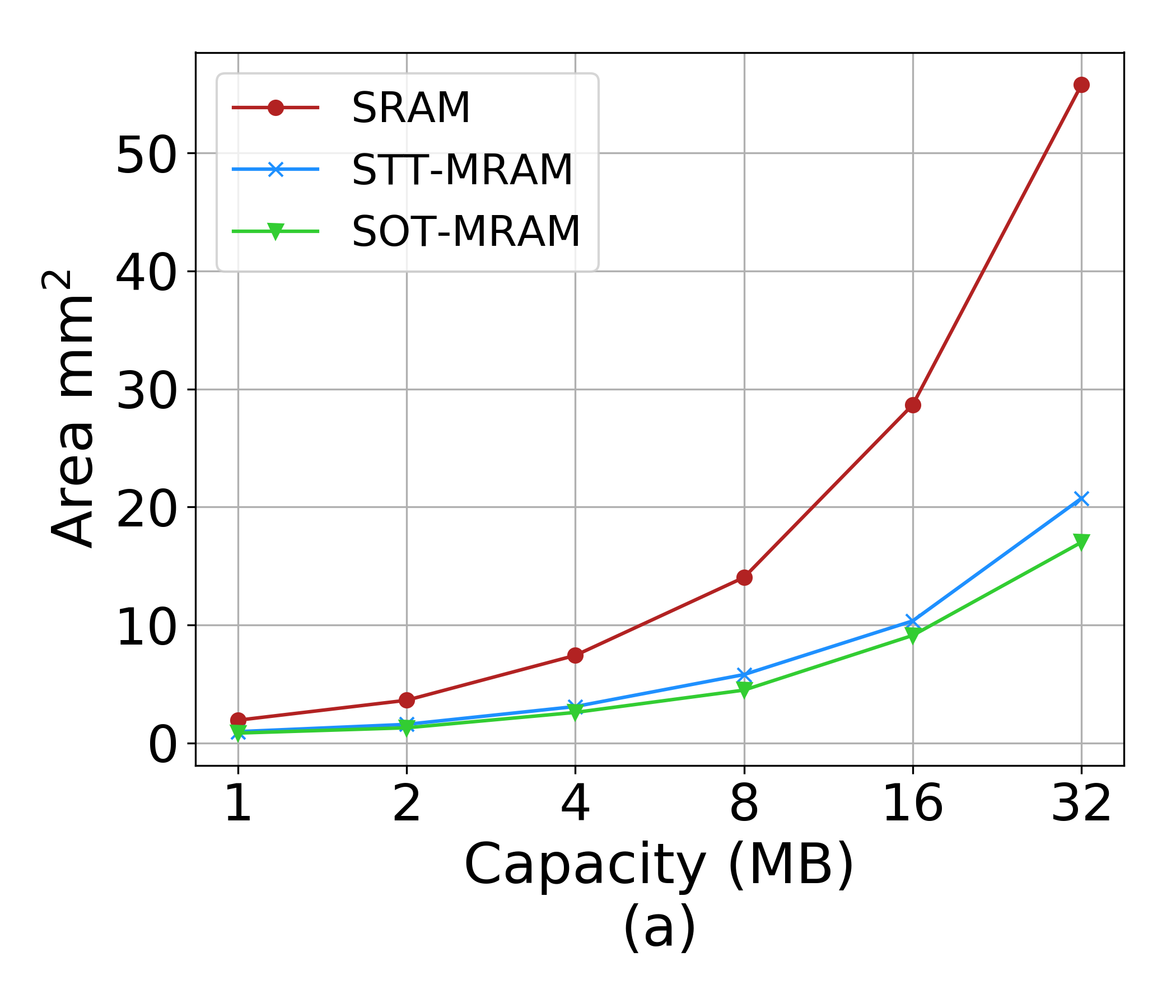}}
  \subfloat{\includegraphics[width=0.32\textwidth]{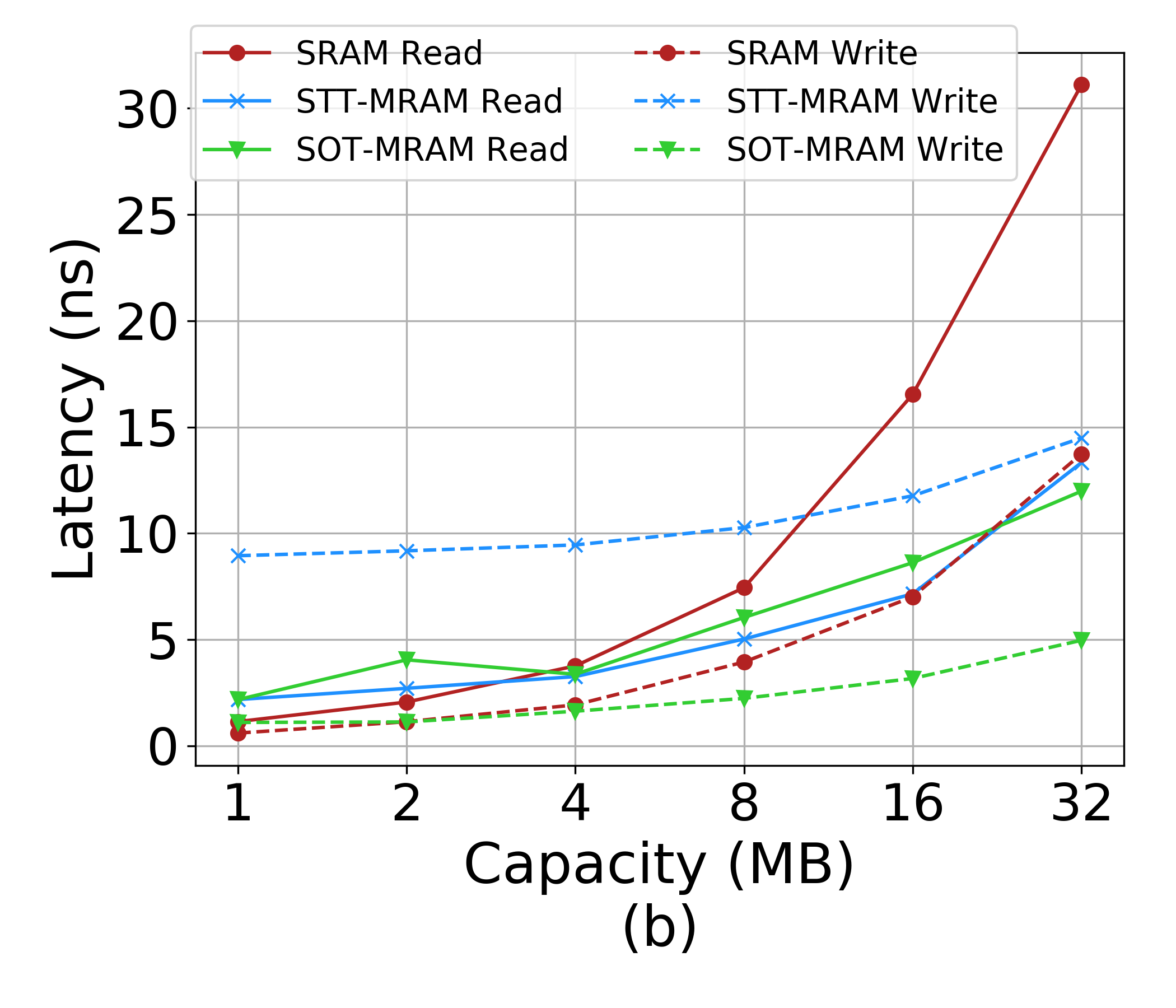}}
  \subfloat{\includegraphics[width=0.32\textwidth]{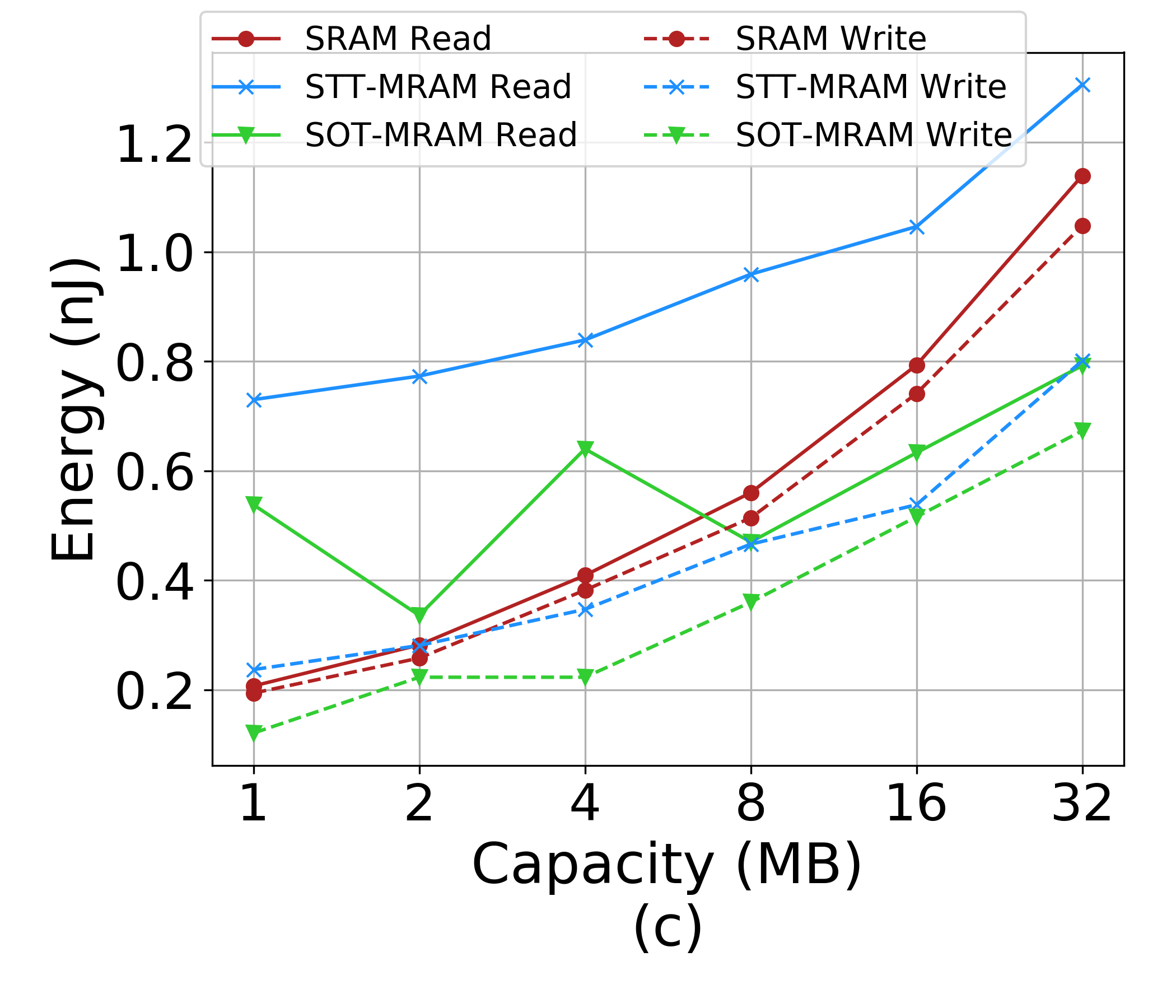}}
 \caption{{Cache capacity scaling results for SRAM, STT-MRAM, and SOT-MRAM for area, latency, and energy metrics}}\label{fig:scaling}
\end{figure*}

As shown in Figure~\ref{fig:trend}, the current trend for NVIDIA GPUs is towards increasing L2 size with each new GPU generation. The most recent high-end NVIDIA GPUs have even up to 6MB L2 cache to further improve performance of the system by reducing costly off-chip memory accesses. However, SRAM has a scalability problem due to its high leakage and large bitcell area, which poses a significant challenge to further continue the current GPU trend. 
To this end, non-volatile memory technologies come to the rescue of future GPU architectures since their PPA scale better as cache capacity increases. Therefore, there is a need for a scalability analysis to project and quantify performance and energy gains that can be achieved by using more scalable memory solutions. 

To this end, we perform a scalability analysis by first comparing SRAM, STT-MRAM, and SOT-MRAM for various cache capacities in terms of area, latency, energy results following the \textit{DeepNVM++} framework methodology as described in Section III. Therefore, each memory technology is optimized for EDAP objective at each cache capacity independently to make a fair comparison among SRAM, STT-MRAM, and SOT-MRAM. Next, we evaluate and show how NVM-based caches behave in terms of performance and energy when compared to conventional SRAM-based caches for deep learning workloads in a scalability analysis. 

\textbf{Area.}
Figure \ref{fig:scaling}(a) demonstrates the impact of higher cell density of MRAMs on the area of caches compared to SRAM. The area difference between SRAM and the MRAM variants grow significantly as the cache capacity increases. The main reason of this difference comes from the bitcell area difference between SRAM and MRAMs as shown in the last row of Table \ref{table:powertable}. Particularly for deeply scaled technology nodes wherein interconnects account for a significant portion of parasitics, bigger bitcells translate to longer wires, bigger buffers, and peripheral logic. Therefore, STT-MRAM and SOT-MRAM caches become more area-efficient when compared to SRAM caches as cache capacity increases.

\textbf{Latency.}
Figure \ref{fig:scaling}(b) shows that for capacities smaller than 3MB SRAM offers lower read latency, whereas both MRAM variants have lower read latency than SRAM beyond 4MB. In terms of write latency, STT-MRAM has always the highest among all memory technologies due to its inherent device characteristic. In contrast, the write latency of SOT-MRAM becomes increasingly smaller than that of SRAM. Moreover, the write latency of SRAM almost matches that of STT-MRAM at 32MB.

\textbf{Energy.}
In terms of read access energy, Figure \ref{fig:scaling}(c) shows that 7MB is a break even point where SOT-MRAM becomes more efficient than SRAM whereas STT-MRAM clearly has the highest read energy among all memories. Regarding write access energy, SOT-MRAM is the most efficient option whereas SRAM consumes the most energy for a write operation beyond 3MB.

\begin{figure*}[!tbp]
  \centering
  \subfloat{\includegraphics[width=0.49\textwidth]{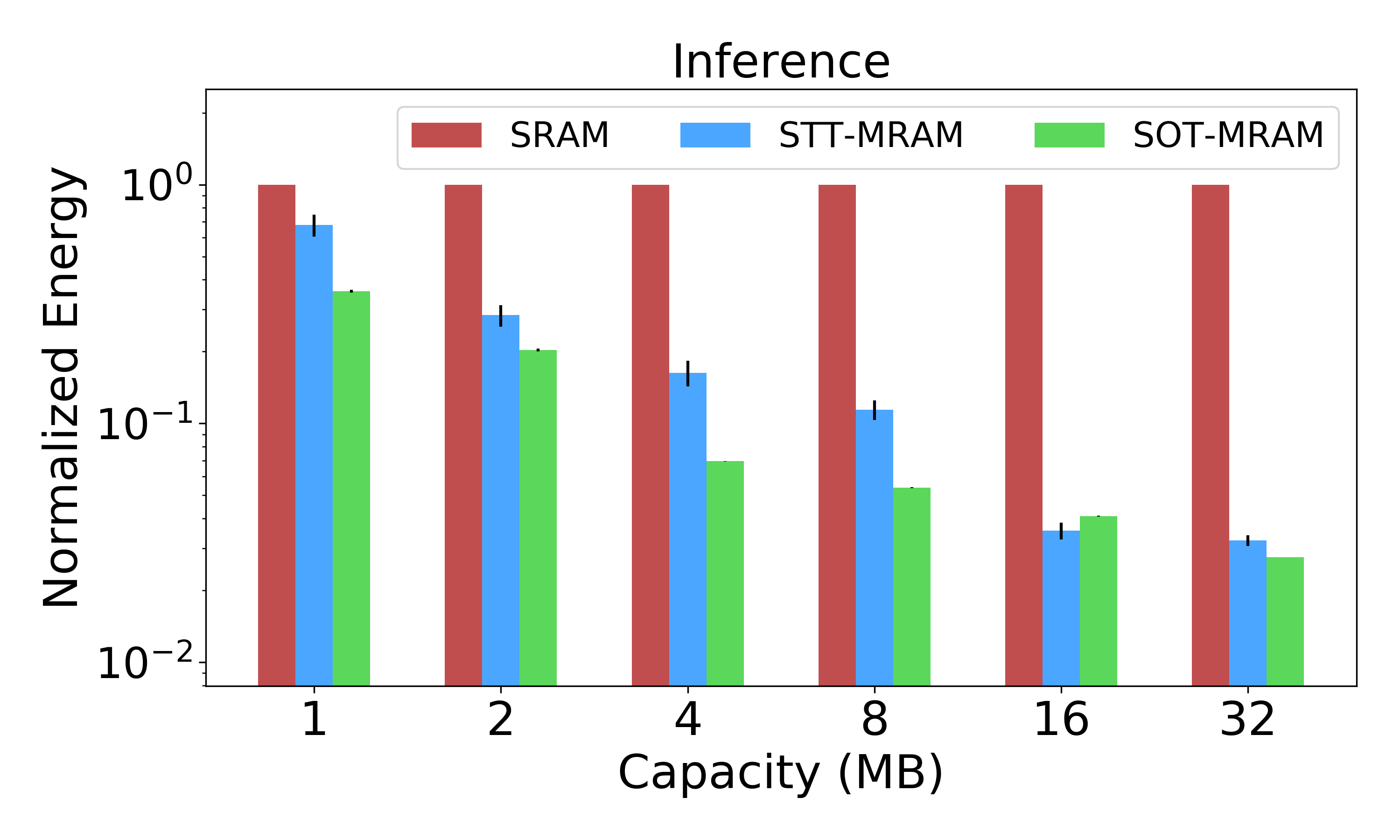}}
  \subfloat{\includegraphics[width=0.49\textwidth]{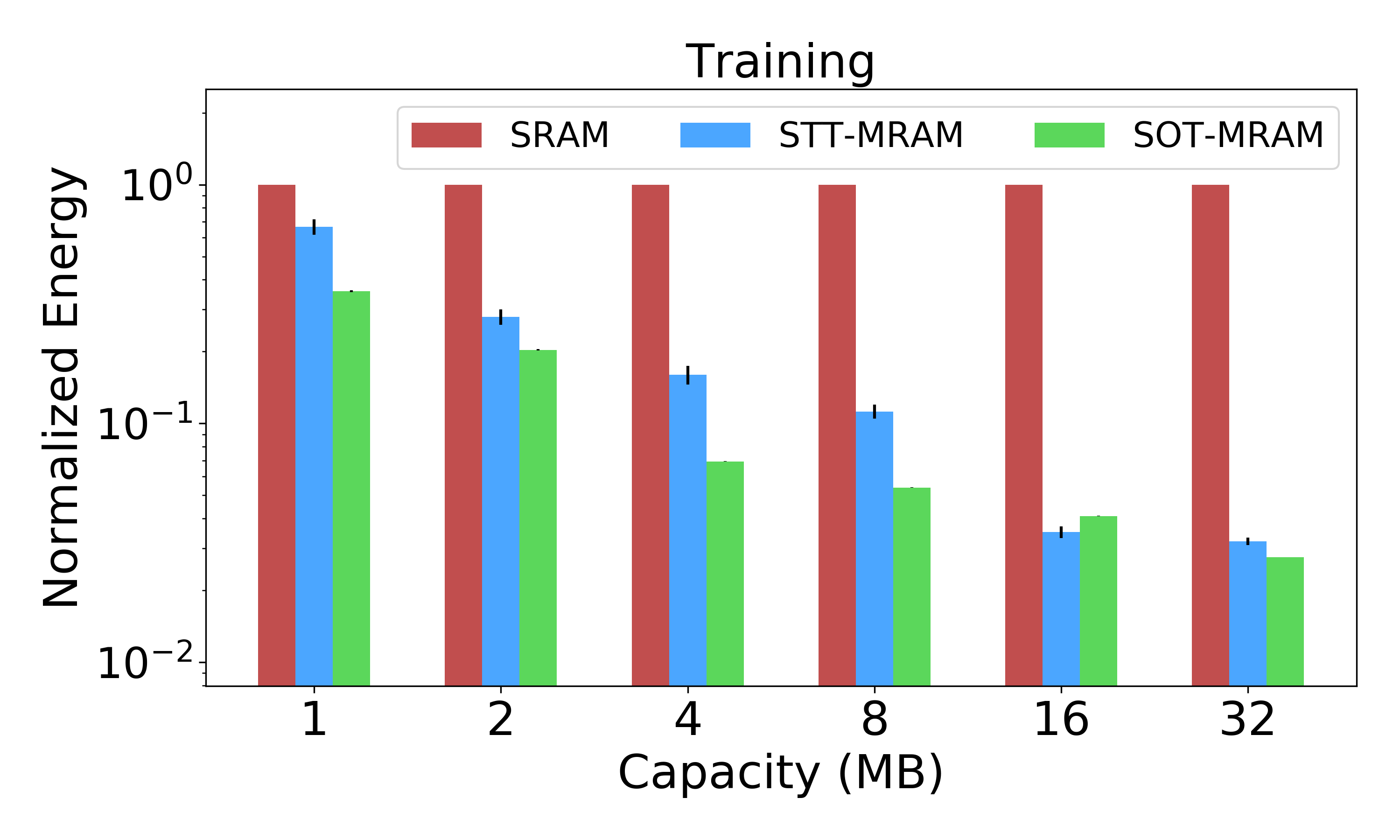}}
  \\
 \subfloat{\includegraphics[width=0.49\textwidth]{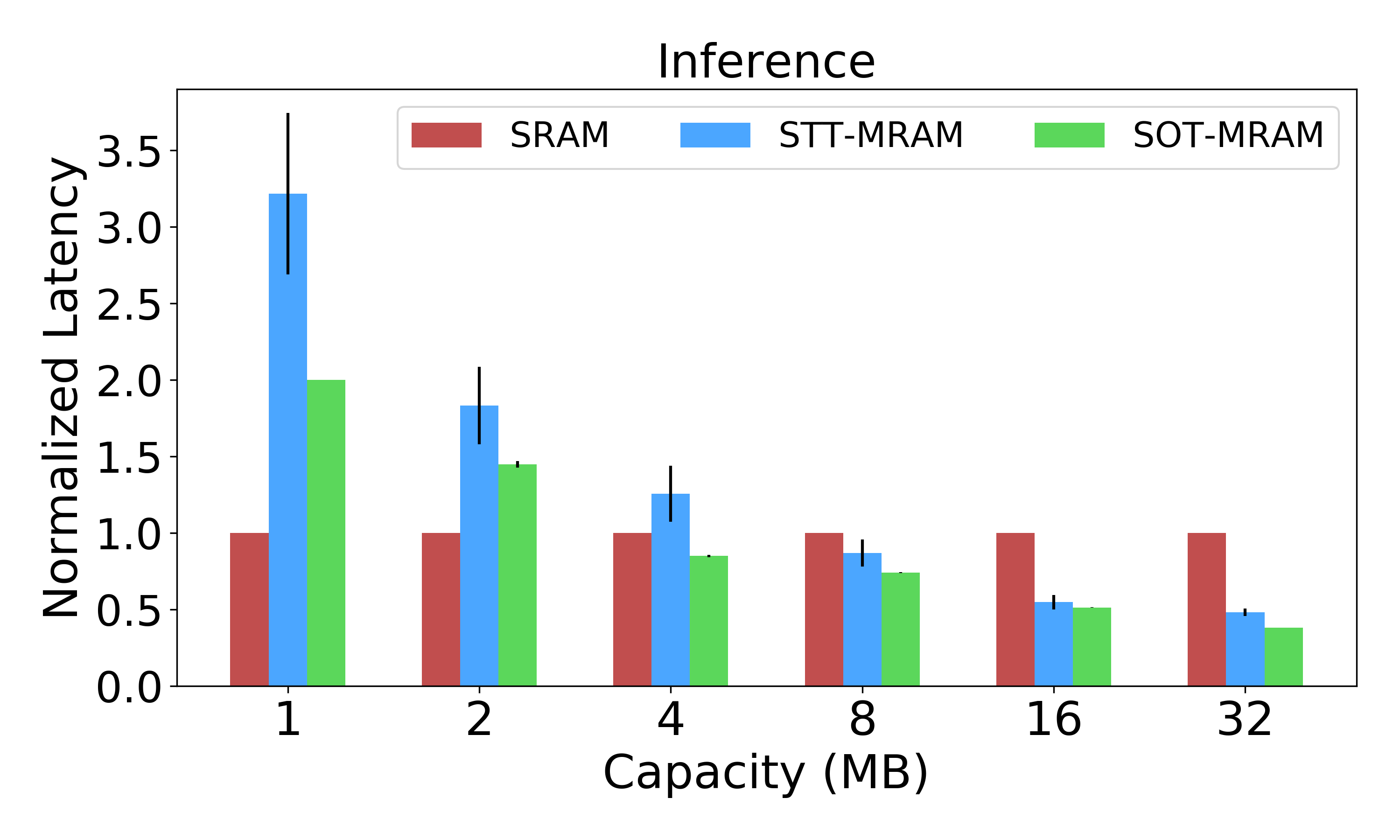}}
  \subfloat{\includegraphics[width=0.49\textwidth]{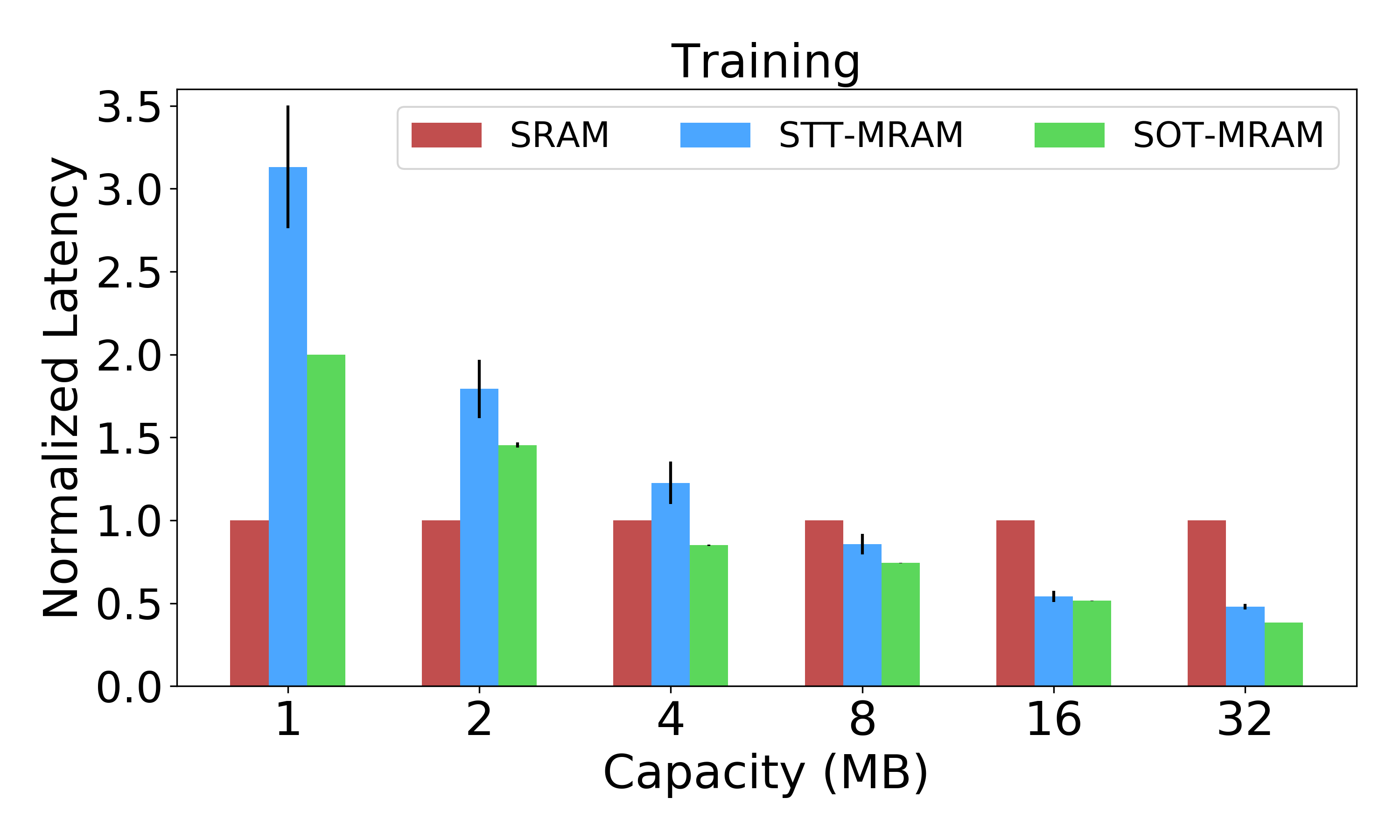}}
 \\
    \subfloat{\includegraphics[width=0.49\textwidth]{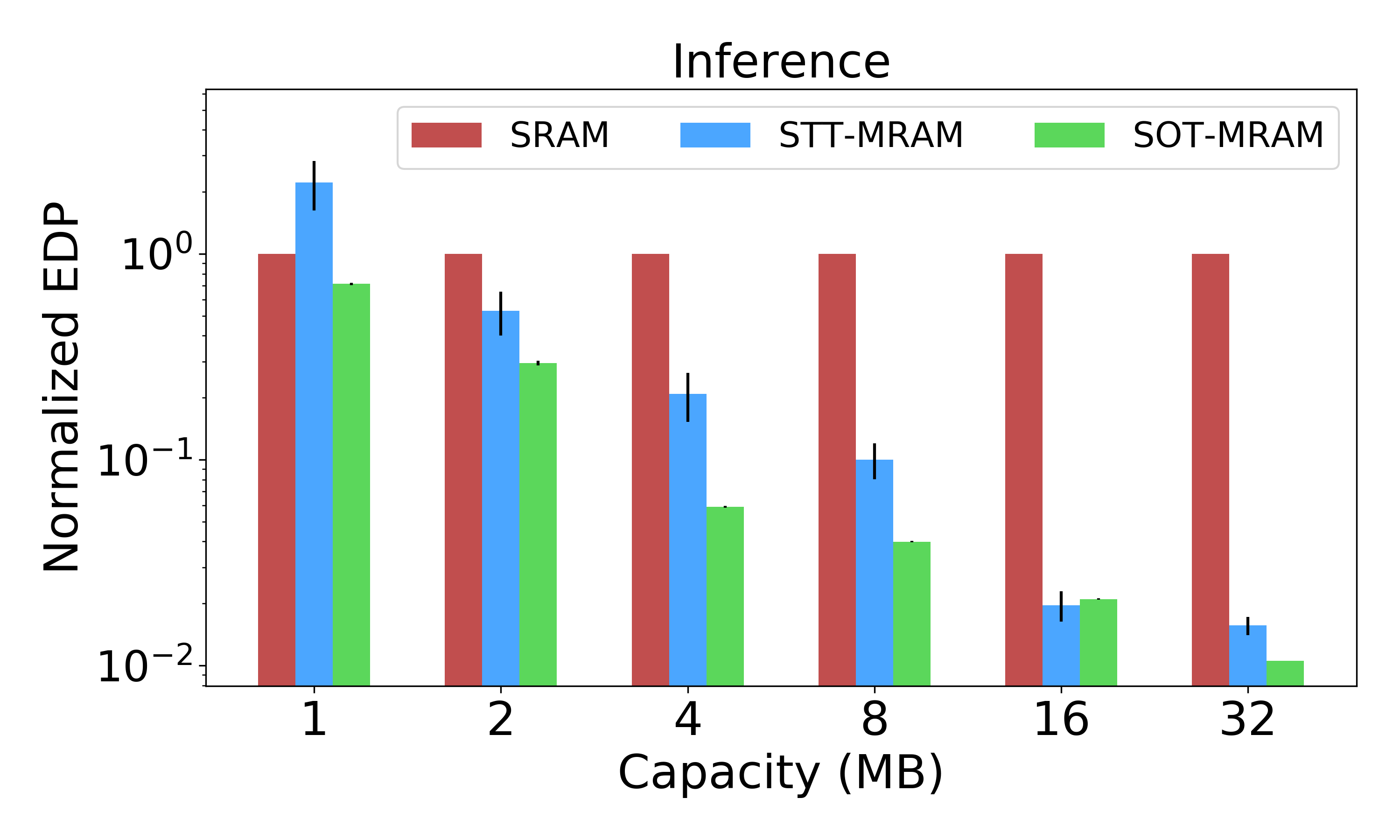}}
      \subfloat{\includegraphics[width=0.49\textwidth]{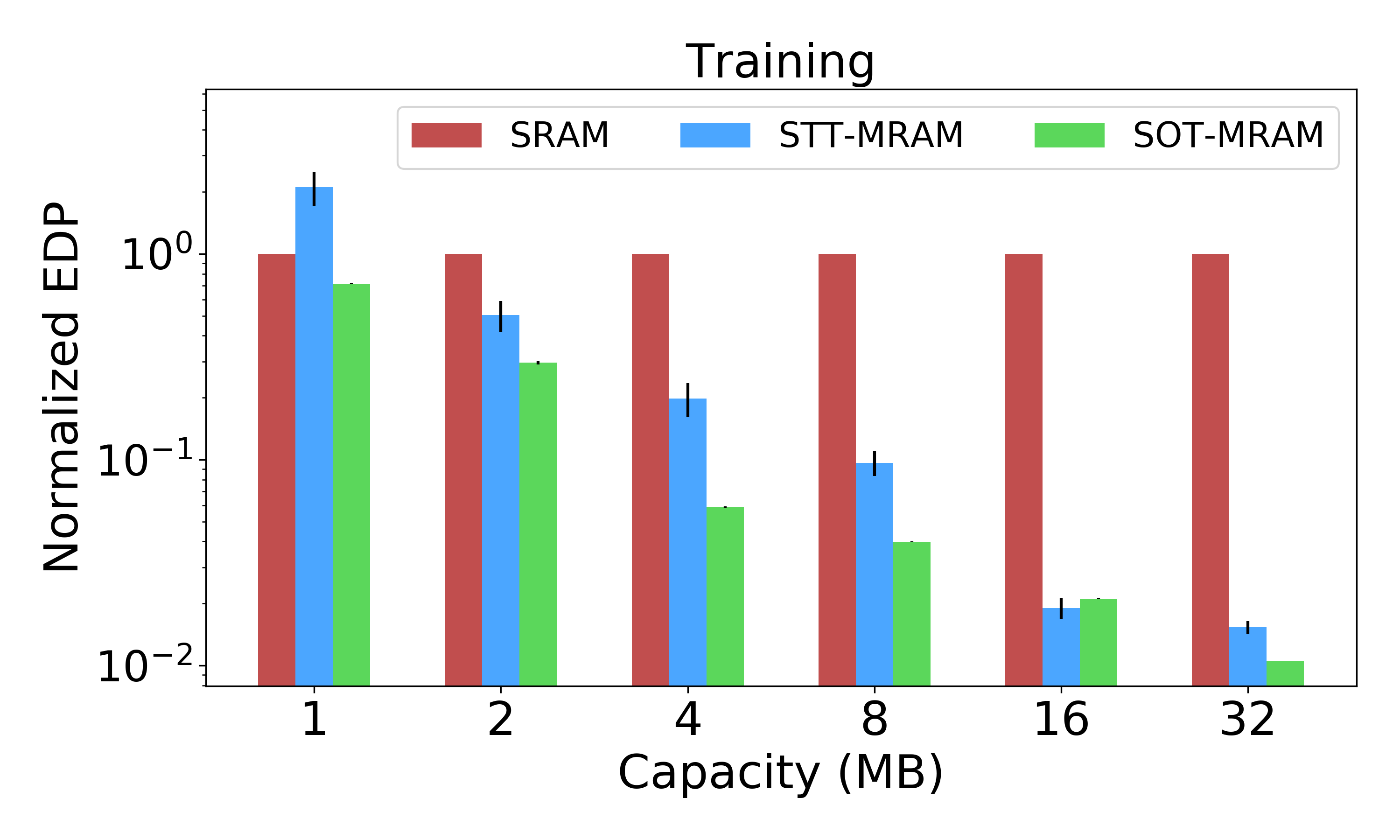}}
\caption{Mean energy (top), latency (middle), and energy-delay product (bottom) across all workloads (lower is better) normalized with respect to SRAM for various cache capacities for inference (left) and training (right) stages. Error bars show standard deviation across workloads.}
\label{fig:norm_inference_edp}
\end{figure*}

Based on these PPA results, we perform a detailed scalability analysis for SRAM, STT-MRAM, and SOT-MRAM. In Figure~\ref{fig:norm_inference_edp}, we show the normalized energy, latency, and EDP results with respect to SRAM for STT-MRAM and SOT-MRAM for various cache capacities. As it can be seen, STT-MRAM and SOT-MRAM provide lower energy and latency results as cache capacity increases.

In terms of energy, STT-MRAM and SOT-MRAM provide lower energy as cache capacity increases. Specifically, STT-MRAM and SOT-MRAM caches achieve up to \textit{$31.2 \times$ and $36.4 \times$ energy reduction} as cache capacity increases, respectively. In terms of latency, STT-MRAM and SOT-MRAM have higher latency results for cache capacities up to 4MB, whereas both MRAM variants have lower latency results when compared to SRAM beyond that point. In more detail, SRAM provide up to \textit{$3.2 \times$ and $2 \times$ latency reduction} for small cache capacities when compared to STT-MRAM and SOT-MRAM, respectively. However, STT-MRAM and SOT-MRAM achieve up to \textit{$2.1 \times$ and $2.6 \times$ latency reduction} as cache capacity increases, respectively. In terms of EDP, we show that STT-MRAM and SOT-MRAM provide up to \textit{$65 \times$ and $95 \times$ EDP reduction} when compared to SRAM, respectively. 

Therefore, we conclude that for latency-critical applications, SRAM-based caches become a more suitable option when compared to MRAM variants for small cache capacities whereas MRAMs provide more energy-efficient solutions. 
Although SRAM provide lower EDP results for smaller cache capacities, STT-MRAM and SOT-MRAM outperform SRAM by orders of magnitude for larger cache capacities due to their better PPA scalability when compared to SRAM. 
These results show that a significant portion of the overall system energy or latency is saved and can be used for additional on-chip resources or capabilities that are not available now.


\section{Discussion}
In this section, we discuss the implications of the results shown in this paper. We also share the potential future directions to guide our community to better explore the use of non-volatile memories for deep learning workloads in different design spaces.

\textbf{Scalability is a major problem for SRAM.} \space\space As we show in Figure~\ref{fig:scaling} and Section IV-C, one of the key challenges for the current GPU architectures is the scalability problem of SRAM due to its significantly high leakage energy and large area when compared to STT-MRAM and SOT-MRAM. We observe that there is a current trend in GPU architectures towards increasing L2 cache capacity and we show that SRAM has significant scalability problems in terms of area, latency, and energy. We show that STT-MRAM and SOT-MRAM have promising solutions for larger cache capacities which can maintain the current trend shown in Figure~\ref{fig:trend} with increasing performance and energy benefits.

\textbf{Implications of dense NVM caches on logic usage.} \space\space Figure~\ref{fig:scaling}(a) shows the area results for SRAM, STT-MRAM, and SOT-MRAM for various cache capacities. We note that STT-MRAM and SOT-MRAM provide increasingly smaller area than SRAM as cache capacity increases. For the same cache capacity, STT-MRAM and SOT-MRAM provide 58\% and 65\% area reduction on average, respectively. Therefore, the remaining whitespace can be utilized by cramming more processing elements, register files, or L2 cache on the die. This analysis is left for future work. 

As CMOS scaling issues limit the affordable improvement of computing systems, our results from device-level simulations to actual GPU profiling show that MRAMs are extremely promising candidates. Particularly, as STT-MRAM and SOT-MRAM fabrication processes become more mature, system-level benefits of STT-MRAM and SOT-MRAM can be maximized, enabling faster and more energy-efficient computation. 

\textbf{Mobile design space exploration for NVM.} \space\space
In this work, we explore the GPU architecture design space to unveil the potential of non-volatile memories for deep learning workloads. Having said that, we note that inference at the edge devices also becomes a common practice for many service providers such as Google~\cite{google_smartreply}, Amazon~\cite{alexa}, and Facebook~\cite{fb} to improve user experience by reducing latency and preserving the private user data on device \cite{ghodsi2020cryptonas}. To this end, Wu \textit{et al}.~\cite{fb} shows that majority of mobile inference for Facebook workloads run on mobile CPUs. Mobile platforms have various resource constraints such as energy, memory, and computing capabilities. Thus, last-level caches of mobile CPUs or hardware accelerators can also be replaced by STT-MRAM and SOT-MRAM to improve performance and energy by reducing leakage energy and costly off-chip memory accesses due to their non-volatility and higher cell density \cite{llc_nvm_intel,hankin2019modernuse,pentecost2019maxnvm,li2019dse}. Therefore, the design space exploration of STT-MRAM and SOT-MRAM for mobile CPUs and hardware accelerators for inference workloads merits further research.

\section{Conclusion}
In this paper, we present the first cross-layer analysis framework to characterize, model, and analyze various NVM technologies in GPU architectures for deep learning workloads. Our novel framework can be used to further explore the feasibility of emerging NVM technologies for DL applications for different design choices such as technology nodes, bitcell models, DL workloads, cache configurations, optimization targets, and target platforms. 

Our results show that in the iso-capacity case, STT-MRAM and SOT-MRAM provide up to \textit{$3.8 \times$ and $4.7 \times$ EDP reduction and $2.4 \times$ and $2.8 \times$ area reduction} when compared to SRAM, respectively. In the iso-area case, STT-MRAM and SOT-MRAM achieve up to \textit{$2 \times$ and $2.3 \times$ EDP reduction} and accommodate \textit{$2.3 \times$ and $3.3 \times$ cache capacity} when compared to SRAM, respectively. Finally, we perform a scalability analysis and show that STT-MRAM and SOT-MRAM outperform their SRAM counterpart by orders of magnitude in terms of energy-delay product for large cache capacities. The newly created energy or latency slack can be used for additional on-chip resources or capabilities that are currently not possible. 

\section*{Acknowledgment}
This research was supported in part by NSF CCF Grant No. 1815899 and CSR Grant No. 1815780.

\ifCLASSOPTIONcaptionsoff
  \newpage
\fi



%

\bibliographystyle{IEEEtran}
\bibliography{main}

\begin{thebibliography}{10}
\providecommand{\url}[1]{#1}
\csname url@samestyle\endcsname
\providecommand{\newblock}{\relax}
\providecommand{\bibinfo}[2]{#2}
\providecommand{\BIBentrySTDinterwordspacing}{\spaceskip=0pt\relax}
\providecommand{\BIBentryALTinterwordstretchfactor}{4}
\providecommand{\BIBentryALTinterwordspacing}{\spaceskip=\fontdimen2\font plus
\BIBentryALTinterwordstretchfactor\fontdimen3\font minus
  \fontdimen4\font\relax}
\providecommand{\BIBforeignlanguage}[2]{{%
\expandafter\ifx\csname l@#1\endcsname\relax
\typeout{** WARNING: IEEEtran.bst: No hyphenation pattern has been}%
\typeout{** loaded for the language `#1'. Using the pattern for}%
\typeout{** the default language instead.}%
\else
\language=\csname l@#1\endcsname
\fi
#2}}
\providecommand{\BIBdecl}{\relax}
\BIBdecl

\bibitem{Wulf:1995:HMW:216585.216588}
\BIBentryALTinterwordspacing
W.~A. Wulf and S.~A. McKee, ``Hitting the memory wall: Implications of the
  obvious,'' \emph{SIGARCH Comput. Archit. News}, vol.~23, no.~1, p. 20–24,
  Mar. 1995. [Online]. Available: \url{https://doi.org/10.1145/216585.216588}
\BIBentrySTDinterwordspacing

\bibitem{dennard}
R.~H. {Dennard}, F.~H. {Gaensslen}, H.~{Yu}, V.~L. {Rideout}, E.~{Bassous}, and
  A.~R. {LeBlanc}, ``Design of ion-implanted mosfet's with very small physical
  dimensions,'' \emph{IEEE Journal of Solid-State Circuits}, vol.~9, no.~5, pp.
  256--268, Oct 9(5):256-268, 1974.

\bibitem{temp_control}
S.~Murali, A.~Mutapcic, D.~Atienza, R.~Gupta, S.~Boyd, L.~Benini, and
  G.~De~Micheli, ``Temperature control of high-performance multi-core platforms
  using convex optimization,'' in \emph{Proceedings of the Conference on
  Design, Automation and Test in Europe}, March 2008, pp. 110--115.

\bibitem{coskun2007temp}
A.~K. {Coskun}, T.~S. {Rosing}, and K.~{Whisnant}, ``Temperature aware task
  scheduling in mpsocs,'' in \emph{2007 Design, Automation Test in Europe
  Conference Exhibition}, 2007, pp. 1--6.

\bibitem{coskun2008temp}
\BIBentryALTinterwordspacing
A.~K. Coskun, T.~S. Rosing, K.~A. Whisnant, and K.~C. Gross, ``Static and
  dynamic temperature-aware scheduling for multiprocessor socs,'' \emph{IEEE
  Trans. Very Large Scale Integr. Syst.}, vol.~16, no.~9, p. 1127–1140, Sep.
  2008. [Online]. Available: \url{https://doi.org/10.1109/TVLSI.2008.2000726}
\BIBentrySTDinterwordspacing

\bibitem{EfficientNet}
\BIBentryALTinterwordspacing
M.~Tan and Q.~Le, ``{E}fficient{N}et: Rethinking model scaling for
  convolutional neural networks,'' ser. Proceedings of Machine Learning
  Research, K.~Chaudhuri and R.~Salakhutdinov, Eds., vol.~97.\hskip 1em plus
  0.5em minus 0.4em\relax Long Beach, California, USA: PMLR, 09--15 Jun 2019,
  pp. 6105--6114. [Online]. Available:
  \url{http://proceedings.mlr.press/v97/tan19a.html}
\BIBentrySTDinterwordspacing

\bibitem{FixRes}
H.~Touvron, A.~Vedaldi, M.~Douze, and H.~J{\'e}gou, ``Fixing the train-test
  resolution discrepancy,'' in \emph{Advances in Neural Information Processing
  Systems (NeurIPS)}, 2019.

\bibitem{EfficientDet}
M.~Tan, R.~Pang, and Q.~V. Le, ``Efficientdet: Scalable and efficient object
  detection,'' \emph{2020 IEEE/CVF Conference on Computer Vision and Pattern
  Recognition (CVPR)}, pp. 10\,778--10\,787, 2020.

\bibitem{ChipPlacement}
A.~Mirhoseini, A.~Goldie, M.~Yazgan, J.~Jiang, E.~Songhori, S.~Wang, Y.-J. Lee,
  E.~Johnson, O.~Pathak, S.~Bae, A.~Nazi, J.~Pak, A.~Tong, K.~Srinivasa,
  W.~Hang, E.~Tuncer, A.~Babu, Q.~V. Le, J.~Laudon, R.~Ho, R.~Carpenter, and
  J.~Dean, ``Chip placement with deep reinforcement learning,'' 2020.

\bibitem{han2016deep}
S.~Han, H.~Mao, and W.~J. Dally, ``Deep compression: Compressing deep neural
  networks with pruning, trained quantization and huffman coding,'' 2016.

\bibitem{ruizhou2018lightnn}
\BIBentryALTinterwordspacing
R.~Ding, Z.~Liu, R.~D.~S. Blanton, and D.~Marculescu, ``Lightening the load
  with highly accurate storage- and energy-efficient lightnns,'' \emph{ACM
  Trans. Reconfigurable Technol. Syst.}, vol.~11, no.~3, Dec. 2018. [Online].
  Available: \url{https://doi.org/10.1145/3270689}
\BIBentrySTDinterwordspacing

\bibitem{Chin2020CVPR}
T.-W. Chin, R.~Ding, C.~Zhang, and D.~Marculescu, ``Towards efficient model
  compression via learned global ranking,'' in \emph{Proceedings of the
  IEEE/CVF Conference on Computer Vision and Pattern Recognition (CVPR)}, June
  2020.

\bibitem{sram_technology}
M.~{Chang}, P.~{Rosenfeld}, S.~{Lu}, and B.~{Jacob}, ``Technology comparison
  for large last-level caches (l3cs): Low-leakage sram, low write-energy
  stt-ram, and refresh-optimized edram,'' in \emph{2013 IEEE 19th International
  Symposium on High Performance Computer Architecture (HPCA)}, 2013, pp.
  143--154.

\bibitem{sram_leakage}
H.~{Homayoun} and A.~{Veidenbaum}, ``Reducing leakage power in peripheral
  circuits of l2 caches,'' in \emph{2007 25th International Conference on
  Computer Design}, 2007, pp. 230--237.

\bibitem{nvm_llc_stt}
W.~{Xu}, H.~{Sun}, X.~{Wang}, Y.~{Chen}, and T.~{Zhang}, ``Design of last-level
  on-chip cache using spin-torque transfer ram (stt ram),'' \emph{IEEE
  Transactions on Very Large Scale Integration (VLSI) Systems}, vol.~19, no.~3,
  pp. 483--493, 2011.

\bibitem{3dmram2008dac}
{Xiangyu Dong}, {Xiaoxia Wu}, {Guangyu Sun}, {Yuan Xie}, H.~{Li}, and {Yiran
  Chen}, ``Circuit and microarchitecture evaluation of 3d stacking magnetic ram
  (mram) as a universal memory replacement,'' in \emph{2008 45th ACM/IEEE
  Design Automation Conference}, 2008, pp. 554--559.

\bibitem{nvidia_trend}
\emph{List of NVIDIA GPUs,
  https://en.wikipedia.org/wiki/List-of-Nvidia-graphics-processing-units}.

\bibitem{DeepNVM}
A.~F. Inci, M.~M. Isgenc, and D.~Marculescu, ``Deepnvm: A framework for
  modeling and analysis of non-volatile memory technologies for deep learning
  applications,'' in \emph{Proceedings of the 23rd Conference on Design,
  Automation and Test in Europe}, ser. DATE '20, 2020, p. 1295–1298.

\bibitem{mtj}
R.~E. Scheuerlein, ``Magneto-resistive ic memory limitations and architecture
  implications,'' in \emph{Seventh Biennial IEEE International Nonvolatile
  Memory Technology Conference. Proceedings (Cat. No.98EX141)}, June 1998, pp.
  47--50.

\bibitem{stt}
W.~{Zhao}, E.~{Belhaire}, Q.~{Mistral}, C.~{Chappert}, V.~{Javerliac},
  B.~{Dieny}, and E.~{Nicolle}, ``Macro-model of spin-transfer torque based
  magnetic tunnel junction device for hybrid magnetic-cmos design,'' in
  \emph{2006 IEEE International Behavioral Modeling and Simulation Workshop},
  Sept 2006, pp. 40--43.

\bibitem{endurance}
J.~J. {Kan}, C.~{Park}, C.~{Ching}, J.~{Ahn}, Y.~{Xie}, M.~{Pakala}, and S.~H.
  {Kang}, ``A study on practically unlimited endurance of stt-mram,''
  \emph{IEEE Transactions on Electron Devices}, vol.~64, no.~9, pp. 3639--3646,
  Sept 64(9):3639-3646, 2017.

\bibitem{stt_high_current}
M.~{Hosomi}, H.~{Yamagishi}, T.~{Yamamoto}, K.~{Bessho}, Y.~{Higo},
  K.~{Yamane}, H.~{Yamada}, M.~{Shoji}, H.~{Hachino}, C.~{Fukumoto},
  H.~{Nagao}, and H.~{Kano}, ``A novel nonvolatile memory with spin torque
  transfer magnetization switching: spin-ram,'' in \emph{IEEE International
  Electron Devices Meeting, 2005. IEDM Technical Digest.}, 2005, pp. 459--462.

\bibitem{survey}
P.~{Chi}, S.~{Li}, {Yuanqing Cheng}, {Yu Lu}, S.~H. {Kang}, and Y.~{Xie},
  ``Architecture design with stt-ram: Opportunities and challenges,'' in
  \emph{2016 21st Asia and South Pacific Design Automation Conference
  (ASP-DAC)}, 2016, pp. 109--114.

\bibitem{yalamanchili2010stt}
M.~{Rasquinha}, D.~{Choudhary}, S.~{Chatterjee}, S.~{Mukhopadhyay}, and
  S.~{Yalamanchili}, ``An energy efficient cache design using spin torque
  transfer (stt) ram,'' in \emph{2010 ACM/IEEE International Symposium on
  Low-Power Electronics and Design (ISLPED)}, 2010, pp. 389--394.

\bibitem{mehdi2016sot}
G.~{Prenat}, K.~{Jabeur}, P.~{Vanhauwaert}, G.~D. {Pendina}, F.~{Oboril},
  R.~{Bishnoi}, M.~{Ebrahimi}, N.~{Lamard}, O.~{Boulle}, K.~{Garello},
  J.~{Langer}, B.~{Ocker}, M.~{Cyrille}, P.~{Gambardella}, M.~{Tahoori}, and
  G.~{Gaudin}, ``Ultra-fast and high-reliability sot-mram: From cache
  replacement to normally-off computing,'' \emph{IEEE Transactions on
  Multi-Scale Computing Systems}, vol.~2, no.~1, pp. 49--60, 2016.

\bibitem{mehdiNVM}
R.~{Bishnoi}, M.~{Ebrahimi}, F.~{Oboril}, and M.~B. {Tahoori}, ``Architectural
  aspects in design and analysis of sot-based memories,'' in \emph{2014 19th
  Asia and South Pacific Design Automation Conference (ASP-DAC)}, 2014, pp.
  700--707.

\bibitem{oboril2015TCAD}
F.~{Oboril}, R.~{Bishnoi}, M.~{Ebrahimi}, and M.~B. {Tahoori}, ``Evaluation of
  hybrid memory technologies using sot-mram for on-chip cache hierarchy,''
  \emph{IEEE Transactions on Computer-Aided Design of Integrated Circuits and
  Systems}, vol.~34, no.~3, pp. 367--380, 2015.

\bibitem{hybridreg}
{Gushu Li}, {Xiaoming Chen}, {Guangyu Sun}, H.~{Hoffmann}, {Yongpan Liu}, {Yu
  Wang}, and {Huazhong Yang}, ``A stt-ram-based low-power hybrid register file
  for gpgpus,'' in \emph{2015 52nd ACM/EDAC/IEEE Design Automation Conference
  (DAC)}, 2015, pp. 1--6.

\bibitem{hybrid_cache_isca09}
\BIBentryALTinterwordspacing
X.~Wu, J.~Li, L.~Zhang, E.~Speight, R.~Rajamony, and Y.~Xie, ``Hybrid cache
  architecture with disparate memory technologies,'' ser. ISCA '09.\hskip 1em
  plus 0.5em minus 0.4em\relax New York, NY, USA: Association for Computing
  Machinery, 2009, p. 34–45. [Online]. Available:
  \url{https://doi.org/10.1145/1555754.1555761}
\BIBentrySTDinterwordspacing

\bibitem{imani2016hybrid}
M.~{Imani}, S.~{Patil}, and T.~{Rosing}, ``Low power data-aware stt-ram based
  hybrid cache architecture,'' in \emph{2016 17th International Symposium on
  Quality Electronic Design (ISQED)}, 2016, pp. 88--94.

\bibitem{beigi20163dhybrid}
\BIBentryALTinterwordspacing
M.~V. Beigi and G.~Memik, ``Tapas: Temperature-aware adaptive placement for 3d
  stacked hybrid caches,'' in \emph{Proceedings of the Second International
  Symposium on Memory Systems}, ser. MEMSYS '16.\hskip 1em plus 0.5em minus
  0.4em\relax New York, NY, USA: Association for Computing Machinery, 2016, p.
  415–426. [Online]. Available: \url{https://doi.org/10.1145/2989081.2989085}
\BIBentrySTDinterwordspacing

\bibitem{nvm_retention}
C.~W. {Smullen}, V.~{Mohan}, A.~{Nigam}, S.~{Gurumurthi}, and M.~R. {Stan},
  ``Relaxing non-volatility for fast and energy-efficient stt-ram caches,'' in
  \emph{2011 IEEE 17th International Symposium on High Performance Computer
  Architecture}, Feb 2011, pp. 50--61.

\bibitem{stt_adaptable_retention}
K.~{Kuan} and T.~{Adegbija}, ``Energy-efficient runtime adaptable l1 stt-ram
  cache design,'' \emph{IEEE Transactions on Computer-Aided Design of
  Integrated Circuits and Systems}, vol.~39, no.~6, pp. 1328--1339, 2020.

\bibitem{cache_revive_retention}
A.~{Jog}, A.~K. {Mishra}, C.~{Xu}, Y.~{Xie}, V.~{Narayanan}, R.~{Iyer}, and
  C.~R. {Das}, ``Cache revive: Architecting volatile stt-ram caches for
  enhanced performance in cmps,'' in \emph{DAC Design Automation Conference
  2012}, 2012, pp. 243--252.

\bibitem{sun2011retention}
Z.~{Sun}, X.~{Bi}, H.~{Li}, W.~{Wong}, Z.~{Ong}, X.~{Zhu}, and W.~{Wu}, ``Multi
  retention level stt-ram cache designs with a dynamic refresh scheme,'' in
  \emph{2011 44th Annual IEEE/ACM International Symposium on Microarchitecture
  (MICRO)}, 2011, pp. 329--338.

\bibitem{nvm_replacement}
J.~{Wang}, X.~{Dong}, and Y.~{Xie}, ``Oap: An obstruction-aware cache
  management policy for stt-ram last-level caches,'' in \emph{2013 Design,
  Automation Test in Europe Conference Exhibition (DATE)}, 2013, pp. 847--852.

\bibitem{3dmram2009hybrid}
G.~{Sun}, X.~{Dong}, Y.~{Xie}, J.~{Li}, and Y.~{Chen}, ``A novel architecture
  of the 3d stacked mram l2 cache for cmps,'' in \emph{2009 IEEE 15th
  International Symposium on High Performance Computer Architecture}, 2009, pp.
  239--249.

\bibitem{imani2016hybridnarrow}
M.~{Imani}, A.~{Rahimi}, Y.~{Kim}, and T.~{Rosing}, ``A low-power hybrid
  magnetic cache architecture exploiting narrow-width values,'' in \emph{2016
  5th Non-Volatile Memory Systems and Applications Symposium (NVMSA)}, 2016,
  pp. 1--6.

\bibitem{nvsim}
\BIBentryALTinterwordspacing
X.~{Dong}, C.~{Xu}, Y.~{Xie}, and N.~P. {Jouppi}, ``Nvsim: A circuit-level
  performance, energy, and area model for emerging nonvolatile memory,''
  \emph{IEEE Transactions on Computer-Aided Design of Integrated Circuits and
  Systems}, vol. 31(7):994-1007, no.~7, pp. 994--1007, Jul. 31(7):994-1007,
  2012. [Online]. Available: \url{https://doi.org/10.1109/TCAD.2012.2185930}
\BIBentrySTDinterwordspacing

\bibitem{stt_model}
J.~{Kim}, A.~{Chen}, B.~{Behin-Aein}, S.~{Kumar}, J.~{Wang}, and C.~H. {Kim},
  ``A technology-agnostic mtj spice model with user-defined dimensions for
  stt-mram scalability studies,'' in \emph{2015 IEEE Custom Integrated Circuits
  Conference (CICC)}, Sept 2015, pp. 1--4.

\bibitem{sot_model}
M.~{Kazemi}, G.~E. {Rowlands}, E.~{Ipek}, R.~A. {Buhrman}, and E.~G.
  {Friedman}, ``Compact model for spin–orbit magnetic tunnel junctions,''
  \emph{IEEE Transactions on Electron Devices}, vol.~63, no.~2, pp. 848--855,
  Feb 63(2):848-855, 2016.

\bibitem{caffe}
\BIBentryALTinterwordspacing
Y.~Jia, E.~Shelhamer, J.~Donahue, S.~Karayev, J.~Long, R.~Girshick,
  S.~Guadarrama, and T.~Darrell, ``Caffe: Convolutional architecture for fast
  feature embedding,'' in \emph{Proceedings of the 22Nd ACM International
  Conference on Multimedia}, ser. MM '14.\hskip 1em plus 0.5em minus
  0.4em\relax New York, NY, USA: ACM, 2014, pp. 675--678. [Online]. Available:
  \url{http://doi.acm.org/10.1145/2647868.2654889}
\BIBentrySTDinterwordspacing

\bibitem{imagenet}
J.~{Deng}, W.~{Dong}, R.~{Socher}, L.~{Li}, {Kai Li}, and {Li Fei-Fei},
  ``Imagenet: A large-scale hierarchical image database,'' in \emph{2009 IEEE
  Conference on Computer Vision and Pattern Recognition}, 2009, pp. 248--255.

\bibitem{gpgpu-sim}
A.~{Bakhoda}, G.~L. {Yuan}, W.~W.~L. {Fung}, H.~{Wong}, and T.~M. {Aamodt},
  ``Analyzing cuda workloads using a detailed gpu simulator,'' in \emph{2009
  IEEE International Symposium on Performance Analysis of Systems and
  Software}, April 2009, pp. 163--174.

\bibitem{kaushik_roy_sot_area}
Y.~{Seo} and K.~{Roy}, ``High-density sot-mram based on shared bitline
  structure,'' \emph{IEEE Transactions on Very Large Scale Integration (VLSI)
  Systems}, vol.~26, no.~8, pp. 1600--1603, Aug 26(8):1600-1603, 2018.

\bibitem{alexnet}
A.~Krizhevsky, I.~Sutskever, and G.~E. Hinton, ``Imagenet classification with
  deep convolutional neural networks,'' in \emph{Proceedings of the 25th
  International Conference on Neural Information Processing Systems - Volume
  1}, ser. NIPS'12.\hskip 1em plus 0.5em minus 0.4em\relax Red Hook, NY, USA:
  Curran Associates Inc., 2012, p. 1097–1105.

\bibitem{googlenet}
C.~{Szegedy}, {Wei Liu}, {Yangqing Jia}, P.~{Sermanet}, S.~{Reed},
  D.~{Anguelov}, D.~{Erhan}, V.~{Vanhoucke}, and A.~{Rabinovich}, ``Going
  deeper with convolutions,'' in \emph{2015 IEEE Conference on Computer Vision
  and Pattern Recognition (CVPR)}, 2015, pp. 1--9.

\bibitem{vgg16}
\BIBentryALTinterwordspacing
K.~Simonyan and A.~Zisserman, ``Very deep convolutional networks for
  large-scale image recognition,'' \emph{CoRR}, vol. abs/1409.1556, 2014.
  [Online]. Available: \url{http://arxiv.org/abs/1409.1556}
\BIBentrySTDinterwordspacing

\bibitem{resnet50}
K.~{He}, X.~{Zhang}, S.~{Ren}, and J.~{Sun}, ``Deep residual learning for image
  recognition,'' in \emph{2016 IEEE Conference on Computer Vision and Pattern
  Recognition (CVPR)}, 2016, pp. 770--778.

\bibitem{squeezenet}
F.~N. Iandola, S.~Han, M.~W. Moskewicz, K.~Ashraf, W.~J. Dally, and K.~Keutzer,
  ``Squeezenet: Alexnet-level accuracy with 50x fewer parameters and <0.5mb
  model size,'' 2016.

\bibitem{nvprof}
\emph{NVIDIA CUDA Profiler,
  https://docs.nvidia.com/cuda/profiler-users-guide/nvprof-overview}.

\bibitem{darknet}
J.~Redmon, ``Darknet: Open source neural networks in c,''
  \url{http://pjreddie.com/darknet/}, 2013--2016.

\bibitem{eyeriss_ssc}
Y.~{Chen}, T.~{Krishna}, J.~S. {Emer}, and V.~{Sze}, ``Eyeriss: An
  energy-efficient reconfigurable accelerator for deep convolutional neural
  networks,'' \emph{IEEE Journal of Solid-State Circuits}, vol.~52, no.~1, pp.
  127--138, Jan 52(1):127-138, 2017.

\bibitem{eyeriss}
\BIBentryALTinterwordspacing
Y.~Chen, J.~Emer, and V.~Sze, ``Eyeriss: A spatial architecture for
  energy-efficient dataflow for convolutional neural networks,'' in
  \emph{Proceedings of the 43rd International Symposium on Computer
  Architecture}.\hskip 1em plus 0.5em minus 0.4em\relax Piscataway, NJ, USA:
  IEEE Press, 2016, pp. 367--379. [Online]. Available:
  \url{https://doi.org/10.1109/ISCA.2016.40}
\BIBentrySTDinterwordspacing

\bibitem{tetris}
\BIBentryALTinterwordspacing
M.~Gao, J.~Pu, X.~Yang, M.~Horowitz, and C.~Kozyrakis, ``Tetris: Scalable and
  efficient neural network acceleration with 3d memory,'' \emph{SIGARCH
  Computer Architecture News}, vol.~45, no.~1, pp. 751--764, 2017. [Online].
  Available: \url{http://doi.acm.org/10.1145/3093337.3037702}
\BIBentrySTDinterwordspacing

\bibitem{boroumand2018movement}
\BIBentryALTinterwordspacing
A.~Boroumand, S.~Ghose, Y.~Kim, R.~Ausavarungnirun, E.~Shiu, R.~Thakur, D.~Kim,
  A.~Kuusela, A.~Knies, P.~Ranganathan, and O.~Mutlu, ``Google workloads for
  consumer devices: Mitigating data movement bottlenecks,'' in
  \emph{Proceedings of the Twenty-Third International Conference on
  Architectural Support for Programming Languages and Operating Systems}, ser.
  ASPLOS '18.\hskip 1em plus 0.5em minus 0.4em\relax New York, NY, USA:
  Association for Computing Machinery, 2018, p. 316–331. [Online]. Available:
  \url{https://doi.org/10.1145/3173162.3173177}
\BIBentrySTDinterwordspacing

\bibitem{donato2018onchipnvm}
M.~{Donato}, B.~{Reagen}, L.~{Pentecost}, U.~{Gupta}, D.~{Brooks}, and
  G.~{Wei}, ``On-chip deep neural network storage with multi-level envm,'' in
  \emph{2018 55th ACM/ESDA/IEEE Design Automation Conference (DAC)}, 2018, pp.
  1--6.

\bibitem{google_smartreply}
\BIBentryALTinterwordspacing
A.~Kannan, K.~Kurach, S.~Ravi, T.~Kaufmann, A.~Tomkins, B.~Miklos, G.~Corrado,
  L.~Luk{\'{a}}cs, M.~Ganea, P.~Young, and V.~Ramavajjala, ``Smart reply:
  Automated response suggestion for email,'' \emph{CoRR}, vol. abs/1606.04870,
  2016. [Online]. Available: \url{http://arxiv.org/abs/1606.04870}
\BIBentrySTDinterwordspacing

\bibitem{alexa}
\BIBentryALTinterwordspacing
G.~Tucker, M.~Wu, M.~Sun, S.~Panchapagesan, G.~Fu, and S.~Vitaladevuni, ``Model
  compression applied to small-footprint keyword spotting,'' in
  \emph{Interspeech 2016}, 2016, pp. 1878--1882. [Online]. Available:
  \url{http://dx.doi.org/10.21437/Interspeech.2016-1393}
\BIBentrySTDinterwordspacing

\bibitem{fb}
C.~{Wu}, D.~{Brooks}, K.~{Chen}, D.~{Chen}, S.~{Choudhury}, M.~{Dukhan},
  K.~{Hazelwood}, E.~{Isaac}, Y.~{Jia}, B.~{Jia}, T.~{Leyvand}, H.~{Lu},
  Y.~{Lu}, L.~{Qiao}, B.~{Reagen}, J.~{Spisak}, F.~{Sun}, A.~{Tulloch},
  P.~{Vajda}, X.~{Wang}, Y.~{Wang}, B.~{Wasti}, Y.~{Wu}, R.~{Xian}, S.~{Yoo},
  and P.~{Zhang}, ``Machine learning at facebook: Understanding inference at
  the edge,'' in \emph{2019 IEEE International Symposium on High Performance
  Computer Architecture (HPCA)}, Feb 2019, pp. 331--344.

\bibitem{ghodsi2020cryptonas}
Z.~Ghodsi, A.~Veldanda, B.~Reagen, and S.~Garg, ``Cryptonas: Private inference
  on a relu budget,'' 2020.

\bibitem{llc_nvm_intel}
K.~{Korgaonkar}, I.~{Bhati}, H.~{Liu}, J.~{Gaur}, S.~{Manipatruni},
  S.~{Subramoney}, T.~{Karnik}, S.~{Swanson}, I.~{Young}, and H.~{Wang},
  ``Density tradeoffs of non-volatile memory as a replacement for sram based
  last level cache,'' in \emph{2018 ACM/IEEE 45th Annual International
  Symposium on Computer Architecture (ISCA)}, 2018, pp. 315--327.

\bibitem{hankin2019modernuse}
A.~{Hankin}, T.~{Shapira}, K.~{Sangaiah}, M.~{Lui}, and M.~{Hempstead},
  ``Evaluation of non-volatile memory based last level cache given modern use
  case behavior,'' in \emph{2019 IEEE International Symposium on Workload
  Characterization (IISWC)}, 2019, pp. 143--154.

\bibitem{pentecost2019maxnvm}
\BIBentryALTinterwordspacing
L.~Pentecost, M.~Donato, B.~Reagen, U.~Gupta, S.~Ma, G.-Y. Wei, and D.~Brooks,
  ``Maxnvm: Maximizing dnn storage density and inference efficiency with sparse
  encoding and error mitigation,'' in \emph{Proceedings of the 52nd Annual
  IEEE/ACM International Symposium on Microarchitecture}, ser. MICRO '52.\hskip
  1em plus 0.5em minus 0.4em\relax New York, NY, USA: Association for Computing
  Machinery, 2019, p. 769–781. [Online]. Available:
  \url{https://doi.org/10.1145/3352460.3358258}
\BIBentrySTDinterwordspacing

\bibitem{li2019dse}
H.~{Li}, M.~{Bhargav}, P.~N. {Whatmough}, and H.~. {Philip Wong}, ``On-chip
  memory technology design space explorations for mobile deep neural network
  accelerators,'' in \emph{2019 56th ACM/IEEE Design Automation Conference
  (DAC)}, 2019, pp. 1--6.

\end{thebibliography}




%

\begin{IEEEbiography}[{\includegraphics[width=1in,height=1.25in,clip,keepaspectratio]{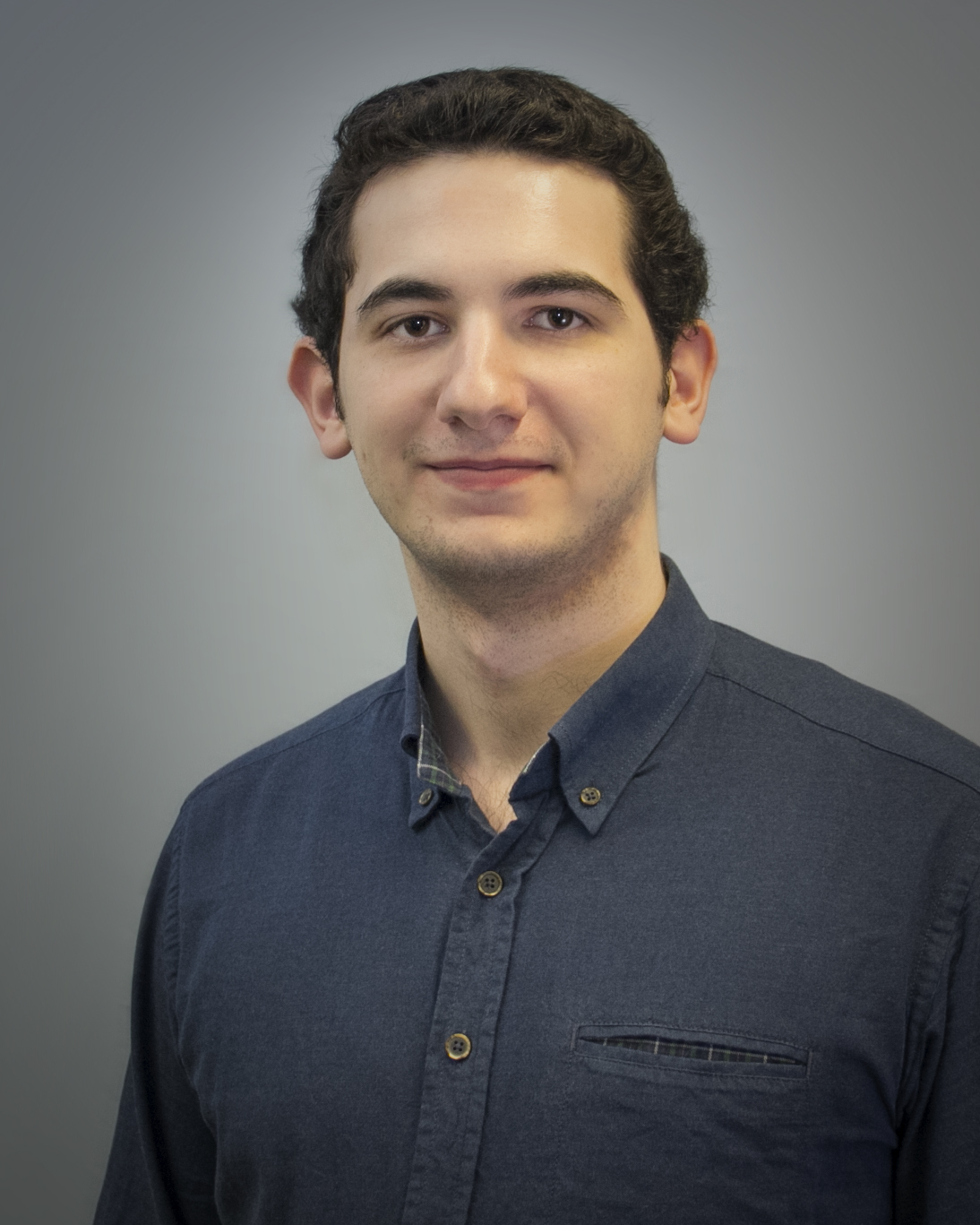}}]{Ahmet Inci} received his B.Sc. degree in Electronics Engineering at Sabanci University, Istanbul, Turkey in 2017. He is currently a Ph.D. candidate at Carnegie Mellon University. His research interests include systems for ML, computer architecture, hardware-efficient deep learning, and HW/ML model co-design. 

\end{IEEEbiography}

\begin{IEEEbiography}[{\includegraphics[width=1in,height=1.25in,clip,keepaspectratio]{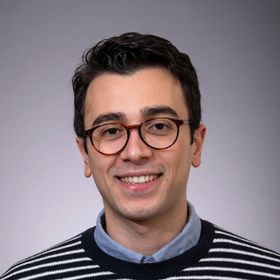}}]{Mehmet Meric Isgenc}
was born in Izmir, Turkey in 1992. He received his B.S in Microelectronics Engineering from Sabanci University in 2014 and his M.Sc and Ph.D. degrees in Electrical and Computer Engineering from Carnegie Mellon University in 2016 and 2019 respectively. He is currently with Apple
Inc. in Cupertino, California. His research interests are low-power SoC accelerator design implementation and automation.
\end{IEEEbiography}

\begin{IEEEbiography}[{\includegraphics[width=1in,height=1.25in,clip,keepaspectratio]{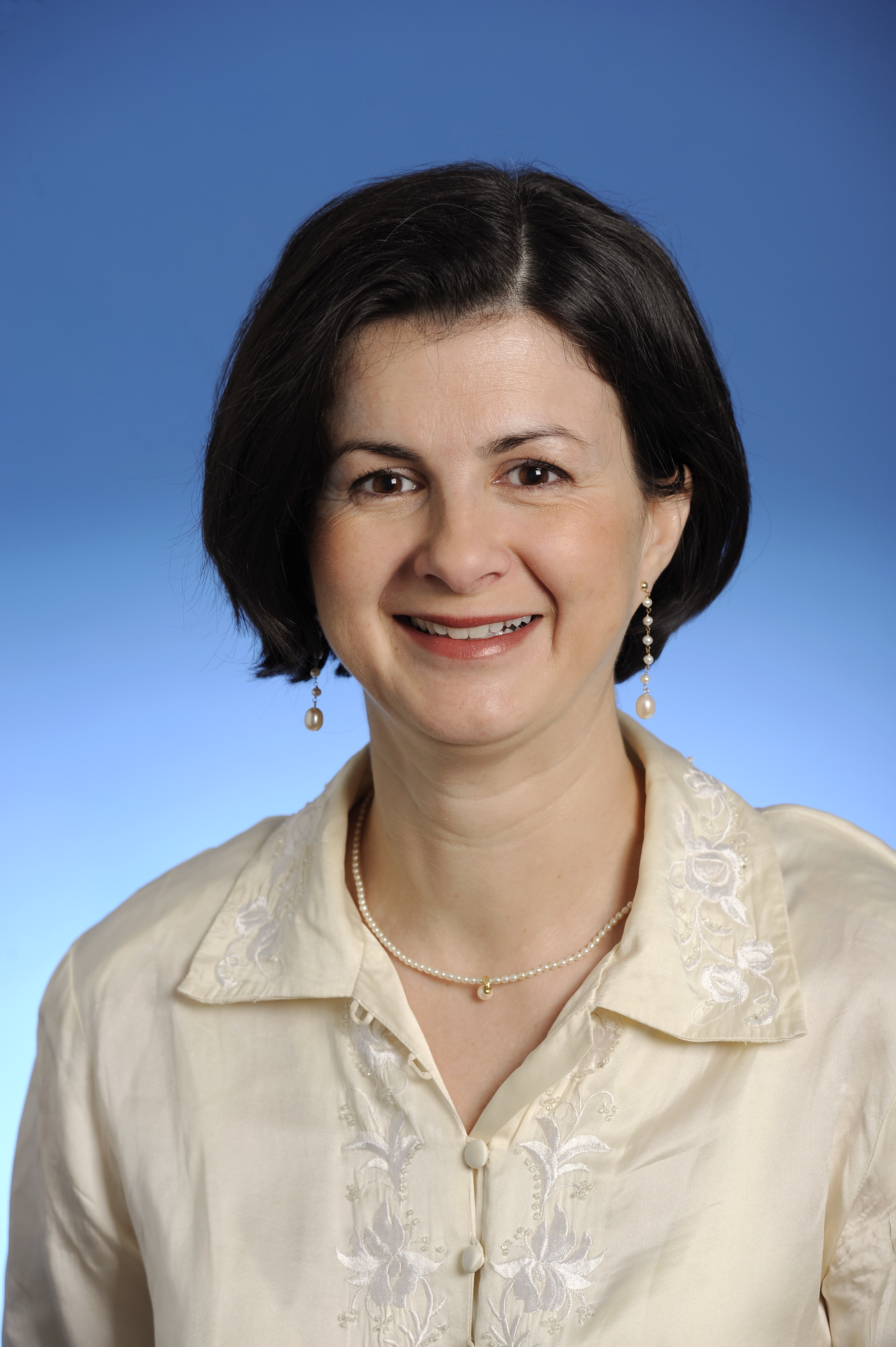}}]{Diana Marculescu}
(Fellow, IEEE) received the
Dipl.Ing. degree in computer science from the Polytechnic University of Bucharest, Bucharest, Romania,
in 1991, and the Ph.D. degree in computer engineering from the University of Southern California, Los
Angeles, CA, USA, in 1998. She is the Department Chair, Cockrell Family Chair for Engineering Leadership \#5, and a Professor, Motorola Regents Chair in Electrical and Computer Engineering \#2, with the University of Texas at Austin. Prior to joining UT Austin in December 2019, she was the David Edward Schramm Professor of Electrical and Computer Engineering, the Founding Director of the College of Engineering Center for Faculty Success (2015–2019) and has served as an Associate Department Head for Academic Affairs in Electrical and Computer Engineering (2014–2018), all with Carnegie Mellon University. Her research interests include energy- and reliability-aware computing, hardware aware machine learning, and computing for sustainability and natural science applications. She was the recipient of the National Science Foundation Faculty Career Award (2000–2004), the ACM SIGDA Technical
Leadership Award (2003), the Carnegie Institute of Technology George Tallman Ladd Research Award (2004), and several best paper awards. She was the IEEE Circuits and Systems Society Distinguished Lecturer (2004–2005) and the Chair of the Association for Computing Machinery (ACM) Special Interest Group on Design Automation (2005–2009). She chaired several conferences and symposia in her area and is currently an Associate Editor for the IEEE TRANSACTIONS ON COMPUTER-AIDED DESIGN OF INTEGRATED CIRCUITS AND SYSTEMS. She was selected as an ELATE Fellow (2013–2014), and is a recipient of an Australian Research Council Future Fellowship (2013–2017), the Marie R. Pistilli Women in EDA Achievement Award (2014), and the Barbara Lazarus Award from Carnegie Mellon University (2018). She is a Fellow of ACM.
\end{IEEEbiography}





\end{document}